\documentclass [12pt, letterpaper]{article}
\usepackage{fullpage}
\usepackage{amsmath}
\usepackage{amssymb}
\usepackage{graphicx}
\usepackage{mathrsfs}
\usepackage{wrapfig}
\usepackage{boxedminipage}
\usepackage{epsfig}
\usepackage{setspace}
\usepackage{subfigure}

\newcommand{\reef}[1]{(\ref{#1})}

\DeclareSymbolFont{AMSb}{U}{msb}{m}{n}
\DeclareMathSymbol{\IN}{\mathbin}{AMSb}{"4E}
\DeclareMathSymbol{\IZ}{\mathbin}{AMSb}{"5A}
\DeclareMathSymbol{\IR}{\mathbin}{AMSb}{"52}
\DeclareMathSymbol{\Q}{\mathbin}{AMSb}{"51}
\DeclareMathSymbol{\II}{\mathbin}{AMSb}{"49}
\DeclareMathSymbol{\IC}{\mathbin}{AMSb}{"43}
\DeclareMathSymbol{\IP}{\mathbin}{AMSb}{"50}
\DeclareMathSymbol{\IH}{\mathbin}{AMSb}{"48}
\DeclareMathSymbol\IA{\mathalpha}{AMSb}{"41}
\DeclareMathSymbol\IS{\mathalpha}{AMSb}{"53}

\def\Q{{\cal Q}}

\begin{document}

\begin{flushright}
\phantom{{\tt arXiv:0709.????}}
\end{flushright}

\bigskip
\bigskip
\bigskip

\begin{center} {\Large \bf Finite Temperature Large $N$ Gauge Theory }
  
  \bigskip

{\Large\bf  with}

\bigskip

{\Large\bf   Quarks in an  External Magnetic Field}

\end{center}

\bigskip \bigskip \bigskip \bigskip

\centerline{\bf Tameem Albash, Veselin Filev, Clifford V. Johnson, Arnab Kundu}

\bigskip
\bigskip
\bigskip

  \centerline{\it Department of Physics and Astronomy }
\centerline{\it University of
Southern California}
\centerline{\it Los Angeles, CA 90089-0484, U.S.A.}

\bigskip

\centerline{\small \tt talbash, filev, johnson1,  akundu  [at] usc.edu}

\bigskip
\bigskip


\begin{abstract} 
\noindent 

Using a ten dimensional dual string background, we study aspects of
the physics of finite temperature large $N$ four dimensional $SU(N)$
gauge theory, focusing on the dynamics of fundamental quarks in the
presence of a background magnetic field. At vanishing temperature and
magnetic field, the theory has ${\cal N}=2$ supersymmetry, and the
quarks are in hypermultiplet representations.  In a previous study,
similar techniques were used to show that the quark dynamics exhibit
spontaneous chiral symmetry breaking. In the present work we begin by
establishing the non--trivial phase structure that results from finite
temperature.  We observe, for example, that above the critical value
of the field that generates a chiral condensate spontaneously, the
meson melting transition disappears, leaving only a discrete spectrum
of mesons at any temperature.  We also compute several thermodynamic
 properties of the  plasma.
\end{abstract}
\newpage \baselineskip=18pt \setcounter{footnote}{0}

\section{Introduction}
\label{sec:introduction}

In recent years, the understanding of the dynamics of a variety of
finite temperature gauge theories at strong coupling has been much
improved by employing several techniques from string theory to capture
the physics.  The framework is that of holographic
\cite{Susskind:1998dq} gauge/gravity duality, in which the physics of
a non--trivial ten dimensional string theory background can be
precisely translated into that of the gauge theory for which the rank
($N$) of the gauge group is
large\cite{Maldacena:1997re,Gubser:1998bc,Witten:1998qj,Witten:1998zw},
while the number ($N_f$) of fundamental flavours of quark is small
compared to $N$ (see ref.\cite{Karch:2002sh}). Many aspects of the
gauge theory, at strong 't Hooft coupling $\lambda=g_{\rm YM}^2N$,
become accessible to computation since the string theory background is
in a regime where the necessary string theory computations are
classical or semi--classical, with geometries that are weakly
curved\cite{Maldacena:1997re} (characteristic radii in the geometry
are set by $\lambda$).
 
These studies are not only of considerable interest in their own
right, but have potential phenomenological applications, since there
are reasons to suspect that they are of relevance to the dynamics of
quark matter in extreme environments such as heavy ion collision
experiments, where the relevant phase seems to be a quark--gluon
plasma.  While the string theory duals of QCD are not known, and will
be certainly difficult to obtain computational control over (the size
$N$ of the gauge group there is small, and the number, $N_f$, of quark
flavours is comparable to $N$) it is expected (and a large and growing
literature of evidence seems to support this -- see below) that there
are certain features of the physics from these accessible models that
may persist to the case of QCD, at least when in a strongly coupled
plasma phase.  Well--studied examples have included various
hydrodynamic properties, such as the ratio of shear viscosity to
entropy, as well as important phenomenological properties of the
interactions between quarks and quark jets with the plasma. Results
from these sorts of computations in the string dual language have
compared remarkably well with QCD phenomenological results from the
heavy ion collision experiments at RHIC (see e.g., refs.
\cite{Shuryak:2003xe,Policastro:2001yc,Kovtun:2003wp,Buchel:2003tz,Liu:2006ug,Gubser:2006bz}),
and have proven to be consistent with and supplementary to results
from the lattice gauge theory approach.

There are many other phenomena of interest to study in a controllable
setting, such as confinement, deconfinement (and the transition
between them), the spectrum and dynamics of baryons and mesons, and
spontaneous chiral symmetry breaking. These models provide a
remarkably clear theoretical laboratory for such physics, as shown for
example in some of the early
work\cite{Babington:2003vm,Kruczenski:2003uq}\ making use of the
understanding of the introduction of fundamental quarks. Some of the
results of these types of studies are also likely to be of interest
for studies of QCD, while others will help map out the possibilities of
what types of physics are available in gauge theories in general, and
guide us toward better control of the QCD physics that we may be able
to probe using gauge/string duals.

This is the spirit of our current paper\footnote{We note that another
  group will present results in this area in a paper to appear
  shortly\cite{johanna}.}. Here, we uncover many new results for a
certain gauge theory at finite temperature and in the presence of a
background external magnetic field, building on work done
recently\cite{Filev:2007gb,Filev:2007qu} on the same theory at zero
temperature.

At vanishing temperature and magnetic field, the large $N$ $SU(N)$
gauge theory has ${\cal N}=2$ supersymmetry, and the quarks are in
hypermultiplet representations. Nevertheless, just as for studies of
the even more artificial ${\cal N}=4$ pure gauge theory, the physics
at finite temperature --- that of a strongly interacting plasma of
quarks and gluons in a variety of phases --- has a lot to teach us
about gauge theory in general, and possibly QCD in particular.

In section two we describe the holographically dual ten--dimensional
geometry and the embedding of the probe D7--brane into it. In section
three we extract the physics of the probe dynamics, using both
analytic and numerical techniques. It is there that we deduce the
phase diagram.  In section four we present our computations of various
thermodynamic properties of the system in various phases, and in
sections five and six we present our computations and results for the
low--lying parts of the spectra of various types of mesons in the
theory. We conclude with a brief discussion in section seven.

\section{The String Background}
Consider the AdS$_5$--Schwarzschild$\times S^5$ solution given by:
\begin{eqnarray}
ds^2/\alpha'&=&-\frac{u^4-b^4}{R^2u^2}dt^2+\frac{u^2}{R^2}d\vec{ x}^2+\frac{R^2u^2}{u^4-b^4}du^2+R^2d\Omega_{5}^2 \ , \\ \mathrm{where}\quad
d\Omega_5^2&=&d\theta^2+\cos^2\theta d\Omega_3^2+\sin^2\theta d\phi^2\nonumber \ ,\\ \mathrm{and}\quad
d\Omega_{3}^2&=&d\psi^2+\cos^2\psi d\beta+\sin^2\theta d\gamma^2 \ .\nonumber
\end{eqnarray}
The dual gauge theory will inherit the time and space coordinates
$t\equiv x^0$ and $\vec{x}\equiv(x^1,x^2,x^3)$ respectively. Also, in
the solution above, $u\in[0,\infty)$ is a radial coordinate on the
asymptotically AdS$_5$ geometry and we are using standard polar
coordinates on the $S^5$. The scale $R$ determines the gauge theory
't Hooft coupling according to $R^2=\alpha^\prime\sqrt{g_{\rm YM}^2 N}$. For
the purpose of our study it will be convenient \cite{Babington:2003vm}
to perform the following change of variables:
\begin{eqnarray} \label{eqt:changeofcoordinates}
&r^2&=\frac{1}{2}(u^2+\sqrt{u^4-b^4})=\rho^2+L^2 \ ,\\ \mathrm{with}\quad
&\rho&=r\cos\theta\ , \,\, L=r\sin\theta\nonumber \ .
\end{eqnarray}
The expression for the metric now takes the form:
\begin{eqnarray}
ds^2/\alpha'&=&-\left(\frac{(4r^4-b^4)^2}{4r^2R^2(4r^4+b^4)}\right)dt^2+\frac{4r^4+b^4}{4R^2r^2}d\vec{x}^2+\frac{R^2}{r^2}(d\rho^2+\rho^2d\Omega_{3}^2+dL^2+L^2d\phi^2)  \ .  \nonumber
\end{eqnarray}
Following ref.~\cite{Karch:2002sh}, we introduce fundamental matter
into the gauge theory by placing D7--brane probes into the dual
supergravity background.  The probe brane is parametrised by the
coordinates $\{x_{0},x_{1},x_{2},x_{3},\rho,\psi,\beta,\gamma\}$ with the
following ansatz for its embedding:
\begin{eqnarray}
\phi\equiv \mathrm{const},\quad L\equiv L(\rho)\nonumber \label{anzatsEmb} \ .
\end{eqnarray}
In order to introduce an external magnetic field, we excite a pure
gauge $B$--field along the $(x^2,x^3)$ directions \cite{Filev:2007gb}:
\begin{equation}
B=Hdx^2\wedge dx^3,
\end{equation}
where $H$ is a real constant. As explained in
ref.~\cite{Filev:2007gb}, while this does not change the
supergravity background, it has a non--trivial effect on the physics of
the probe, which is our focus. To study the effects on the probe, let
us consider the general (Abelian) DBI action:
\begin{eqnarray}
S_{DBI}=- N_f  T_{D7} \int\limits_{{\cal M}_{8}}d^{8}\xi \ \mathrm{det} ^{1/2}(P[G_{ab}+B_{ab}]+2\pi\alpha' F_{ab})\ , \label{DBI}
\end{eqnarray}
where $T_{D7}=\mu_7 / g_s = [(2\pi)^7\alpha'^4 g_s]^{-1}$ is the
D7--brane tension, $P[G_{ab}]$ and $P[B_{ab}]$ are the induced metric
and induced $B$--field on the D7--branes' world--volume, $F_{ab}$ is
the world--volume gauge field, and $N_f=1$ here. It was shown in
ref.~\cite{Filev:2007gb} that, for the AdS$_5\times S^5$ geometry, we
can consistently set the gauge field $F_{ab}$ to zero to leading order
in $\alpha'$, and the same argument applies to the finite temperature
case considered here. The resulting Lagrangian is:
\begin{equation}
{\cal L}=-\rho^3\left(1-\frac{b^8}{16 \left(\rho^2+L(\rho)^2\right)^4}\right) \left\{1+\frac{16 H^2 \left(\rho^2+L(\rho)^2\right)^2 R^4}{\left(b^4+4 \left(\rho^2+L(\rho)^2\right)^2\right)^2}\right\}^{\frac12} 
   \sqrt{1+L'(\rho)^2} \ .
 \end{equation}
For large $\rho \gg b$, the Lagrangian asymptotes to:
\begin{equation}
{\cal L}\approx-\rho^3\sqrt{1+L'(\rho)^2} \ ,
\end{equation}
which suggests the following asymptotic behavior for the embedding
function $L(\rho)$:
\begin{equation}
L(\rho)=m+\frac{c}{\rho^2}+\dots \ ,
\label{asymptote}
\end{equation}
where the parameters $m$ (the asymptotic separation of the D7 and
D3--branes) and $c$ (the degree of transverse bending of the D7--brane
in the $(\rho,\phi)$ plane) are related to the bare quark mass
$m_{q}=m/2\pi\alpha'$ and the fermionic condensate
$\langle\bar\psi\psi\rangle\propto -c$ respectively
\cite{Kruczenski:2003uq} (this calculation is repeated in
appendix~\ref{appendix:condensate}). It was shown in
ref.~\cite{Filev:2007gb} that the presence of the external magnetic
field spontaneously breaks the chiral symmetry of the dual gauge
theory (it generates a non--zero $\langle\bar\psi\psi\rangle$ at zero
$m$). However\cite{Babington:2003vm}, the effect of the finite
temperature is to melt the mesons and restore the chiral symmetry at
zero bare quark mass. Therefore, we have two competing processes
depending on the magnitudes of the magnetic field $H$ and the
temperature $T=b/\pi R^2$.  This suggests an interesting two
dimensional phase diagram for the system, which we shall study in
detail later.

To proceed, it is convenient to define the following dimensionless
parameters:
 \begin{eqnarray}
 \tilde\rho&=&\frac{\rho}{b} \ , ~~~\eta=\frac{R^2}{b^2}H \ , \quad {\tilde m}=\frac{m}{b}\ ,\\
 \tilde L(\tilde\rho)&=&\frac{L(b\tilde\rho)}{b}=\tilde m+\frac{\tilde c}{\tilde \rho^2}+\dots \ .\nonumber
 \label{formula1}
 \end{eqnarray}
This leads to the Lagrangian:
\begin{equation}
\tilde{\cal L}=-\tilde\rho^3\left(1-\frac{1}{16 \left(\tilde\rho^2+\tilde L(\tilde\rho)^2\right)^4}\right) \left\{1+\frac{16\left(\tilde\rho^2+\tilde L(\tilde\rho)^2\right)^2 \eta^2}{\left(1+4 \left(\tilde\rho^2+\tilde L(\tilde\rho)^2\right)^2\right)^2}\right\}^{\frac12}\sqrt{1+\tilde L'(\tilde\rho)^2} \ .
\label{Lagrangian}
\end{equation}
For small values of $\eta$, the analysis of the second order,
non--linear differential equation for $\tilde L(\tilde\rho)$ derived
from equation \reef{Lagrangian} follows closely that performed in
refs.~\cite{Babington:2003vm,Mateos:2006nu,Albash:2006ew}. The
solutions split into two classes: the first class are solutions
corresponding to embeddings that wrap a shrinking $S^3$ in the $S^5$
part of the geometry and (when the $S^3$ vanishes) closes at some
finite radial distance $r$ above the black hole's horizon, which is
located at $r=b/\sqrt{2}$.  These embeddings are referred to as
`Minkowski' embeddings.  The second class of solutions correspond to
embeddings falling into the black hole, since the $S^1$ of the
Euclidean section, on which the D7--branes are wrapped, shrinks away
there.  These embeddings are referred to as `black hole' embeddings.
There is also a critical embedding separating the two classes of
solutions which has a conical singularity at the horizon, where the
$S^3$ wrapped by the D7--brane shrinks to zero size, along with the
$S^1$. If one calculates the free energy of the embeddings, one can
show \cite{Babington:2003vm,Mateos:2006nu,Albash:2006ew} that it is a
multi--valued function of the asymptotic separation $m$, which amounts
to a first order phase transition of the system (giving a jump in the
condensate) for some critical bare quark mass $m_{\rm cr}$. (For fixed
mass, we may instead consider this to be a critical temperature.) We
show in this paper that the effect of the magnetic field is to
decrease this critical mass, and, at some critical magnitude of the
parameter $\eta_{\rm cr}$, the critical mass drops to zero.  For $\eta >
\eta_{\rm cr}$ the phase transition disappears, and only the Minkowski
embeddings are stable states in the dual gauge theory, possessing a
discrete spectrum of states corresponding to quarks and anti--quarks
bound into mesons. Furthermore, at zero bare quark mass, we have a
non--zero condensate and the chiral symmetry is spontaneously broken.
 \section{Properties of the Solution}

 \subsection{Exact Results at Large Mass} 
 It is instructive to first study the properties of the solution for
 $\tilde m \gg 1$. This approximation holds for finite temperature,
 weak magnetic field, and large bare quark mass $m$, or, equivalently,
 finite bare quark mass $m$, low temperature, and weak magnetic field.
 
 In order to analyze the case $\tilde m\gg 1$, let us write $\tilde
 L(\tilde{\rho})=\tilde a +\zeta(\tilde{\rho})$ for $\tilde{a} \gg 1$
 and linearize the equation of motion derived from equation
 \reef{Lagrangian}, while leaving only the first two leading terms in
 $(\rho^2+\tilde m^2)^{-1}$. The result is:
\begin{equation}
\partial_{\tilde\rho}(\tilde\rho^3\zeta')-\frac{2\eta^2}{(\tilde m^2+\tilde\rho^2)^3}\tilde m+\frac{2(\eta^2+1)^2-1}{2(\tilde m^2+\tilde\rho^2)^5}\tilde m+O(\zeta)=0 \ .
\label{bigm}
\end{equation}
 Ignoring the $O(\zeta)$ terms in equation \reef{bigm}, the general solution takes the form:
 \begin{equation}
 \zeta(\tilde \rho)=-\frac{\eta^2}{4\rho^2(\tilde m^2+\tilde\rho^2)}\tilde m+\frac{2(\eta^2+1)^2-1}{96\tilde\rho^2(\tilde m^2+\tilde\rho^2)^3}\tilde m \ ,
 \label{weakL}
 \end{equation}
where we have taken $\zeta'(0) = \zeta(0) = 0$.  By studying the asymptotic behavior of this solution, we can extract the following:
\begin{eqnarray}
\tilde{m} &=& \tilde{a} - \frac{\eta^2}{4 \tilde{a}^3} + \frac{1+ 4 \eta^2 +2 \eta^4}{32 \tilde{a}^7} + O\left(\frac{1}{\tilde{a}^{7}} \right) \nonumber \ , \\
\tilde{c} &=& \frac{\eta^2}{4 \tilde{a}} - \frac{1+ 4\eta^2 +2 \eta^4}{96 \tilde{a}^5} + O\left(\frac{1}{\tilde{a}^7} \right)\ .
\end{eqnarray}
By inverting the expression for $\tilde{m}$, we can express $\tilde{c}$ in terms of $\tilde{m}$:
\begin{eqnarray}
\tilde{c} &=& \frac{\eta^2}{4 \tilde{m}} - \frac{1+ 4 \eta^2 + 8 \eta^4}{96 \tilde{m}^5} + O \left(\frac{1}{\tilde{m}^7} \right) \label{weakcond} \ .
\end{eqnarray}
Finally, after going back to dimensionful parameters, we can see that the theory has developed a fermionic  condensate:
\begin{equation}
\langle\bar\psi\psi\rangle\propto-c=-\frac{R^4}{4m}H^2+\frac{b^8 + 4 b^4 R^4 H^2 + 8 R^8 H^4}{96m^5} \ .
\label{weakc}
\end{equation}
The results of the above analysis can be trusted only for finite bare
quark mass and sufficiently low temperature and weak magnetic field.
As can be expected, the physically interesting properties of the
system should be described by the full non--linear equation of motion
of the D7--brane. To explore these we need to use numerical
techniques.
 %
 \subsection{Numerical Analysis}
\begin{figure}[ht] 
   \centering
   \includegraphics[width=9cm]{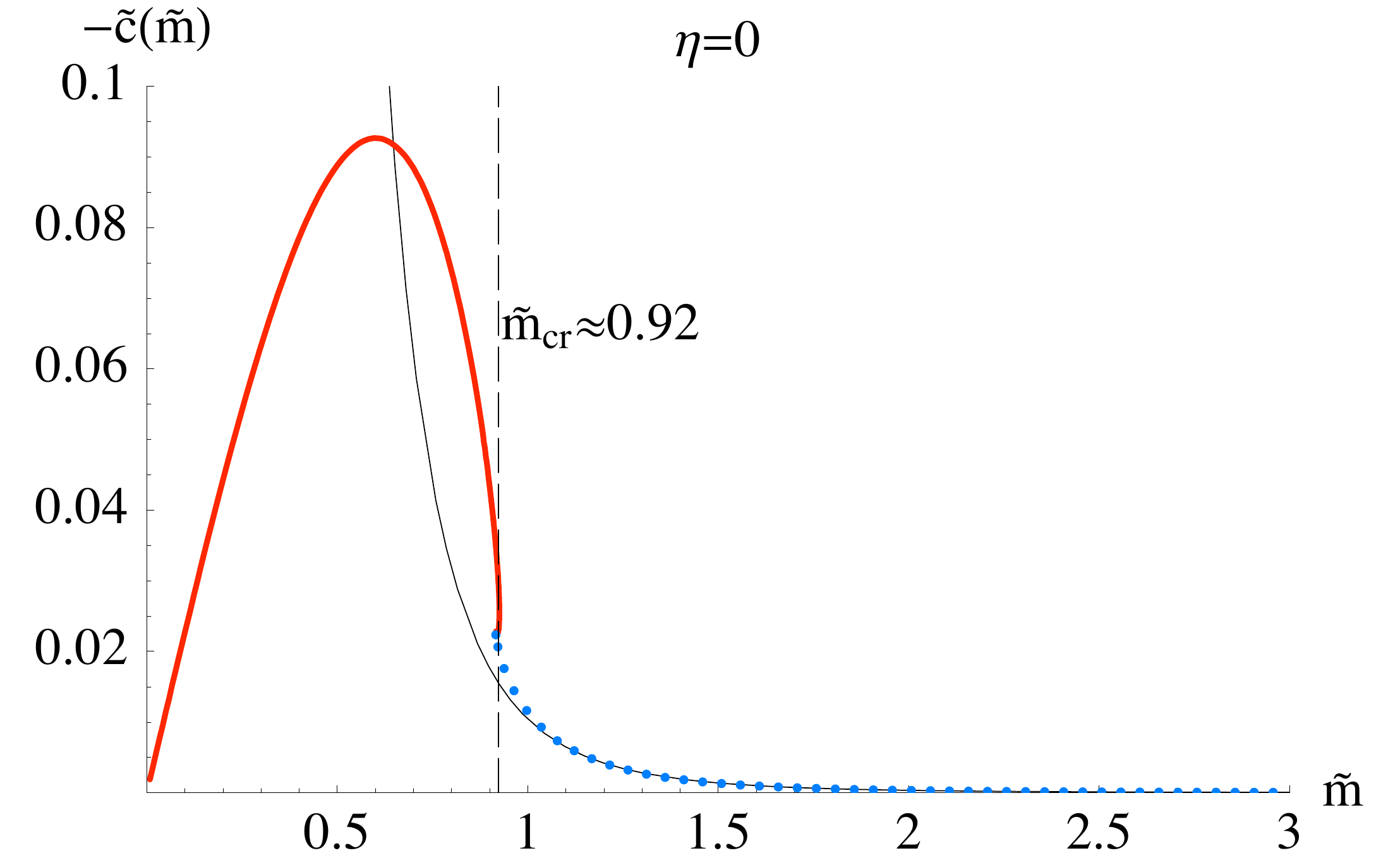} 
   \caption{\small The solid curve starting far left (red) represents solutions falling into the black hole, the dotted  (blue) curve represents solutions with shrinking $S^3$. The vertical dashed line corresponds to the critical value of $\tilde m$ at which the first order phase transition takes place. The solid black curve dropping sharply from  above is the function derived in  equation \reef{weakcond}, corresponding to the large mass limit.}
   \label{fig:1}
\end{figure}

\begin{figure}[ht] 
   \centering
   \includegraphics[width=9cm]{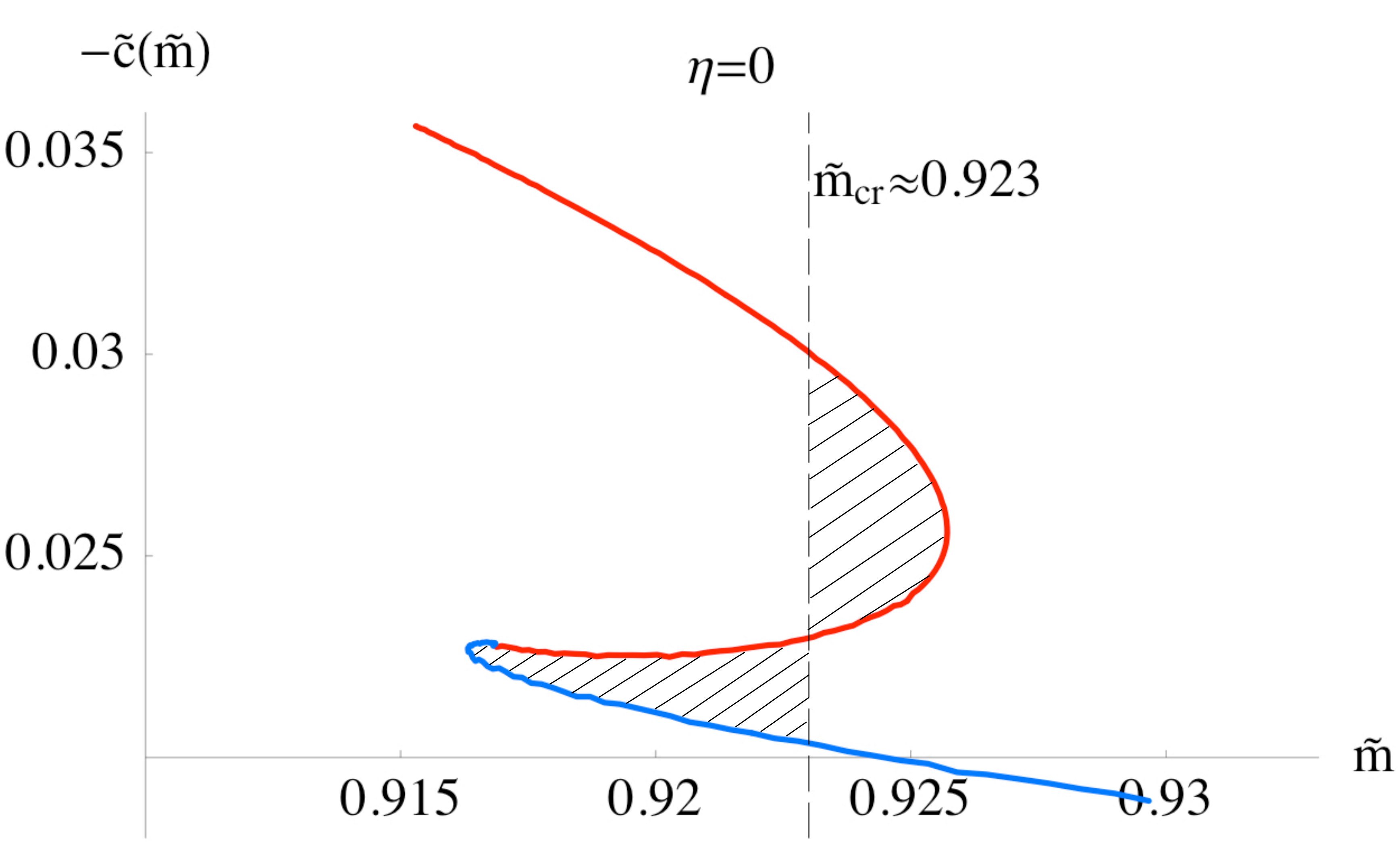} 
   \caption{\small The area below the $(-{\tilde c}, {\tilde m})$ curve has the interpretation of the free energy of the D7--brane; thus the phase transition pattern obeys the ``equal area law''--- the area of the shaded regions is equal.}
   \label{fig:2}
\end{figure}

 We solve the differential equation derived from equation
 \reef{Lagrangian} numerically using Mathematica. It is convenient to
 use infrared initial conditions \cite{Albash:2006ew,Albash:2006bs}.
 For the Minkowski embeddings, based on symmetry arguments, the
 appropriate initial conditions are:
\begin{equation}
\tilde L(\tilde\rho)|_{\tilde\rho=0}=L_{\rm in}, \quad \tilde L'(\tilde\rho)|_{\tilde\rho=0}=0 \ .
\end{equation}
For the black hole embeddings, the following initial conditions:
\begin{equation}
\tilde L(\tilde\rho)|_{\rm e.h.}=\tilde L_{\rm in}, \quad \tilde L'(\tilde\rho)|_{\rm e.h.}=\left.\frac{\tilde L}{\tilde\rho}\right|_{\rm e.h.} \ ,
\end{equation}
ensure regularity of the solution at the event horizon. After solving
numerically for $\tilde L(\tilde\rho)$ for fixed value of the
parameter $\eta$, we expand the solution at some numerically large
$\tilde\rho_{\rm max}$, and, using equation \reef{asymptote}, we generate
the plot of $-\tilde c$ vs $\tilde m$. It is instructive to begin our
analysis by revisiting the case with no magnetic field ($\eta=0)$,
familiar from
refs.\cite{Babington:2003vm,Mateos:2006nu,Albash:2006ew}. The
corresponding plot for this case is presented in figure \ref{fig:1}.
Also in the figure is a plot of the large mass analytic result of
equation~\reef{weakcond}, shown as the thin black curve in the figure,
descending sharply downwards from above; it can be seen that it is
indeed a good approximation for $\tilde m> \tilde m_{\rm cr}$.  Before
we proceed with the more general case of non--zero magnetic field, we
review the techniques employed in ref.~\cite{Albash:2006ew} to
determine the critical value of $\tilde m$. In figure~\ref{fig:2}, we
have presented the region of the phase transition considerably
magnified.
Near the critical value $\tilde m_{\rm cr}$, the condensate $\tilde c$ is
a multi-valued function of $\tilde m$, and we have three competing
phases.  The parameter $\tilde c$ is known\cite{Kruczenski:2003uq} to
be proportional to the first derivative of the free energy of the
D7--brane, and therefore the area below the curve of the $-\tilde c$
vs $\tilde m$ plot is proportional to the free energy of the brane.
Thus, the phase transition happens where the two shaded regions in
figure \ref{fig:2} have equal areas; furthermore, for $\tilde m <
\tilde m_{\rm cr}$, the upper--most branch of the curve corresponds to the
stable phase, and the lower--most branch of the curve corresponds to a
meta--stable phase.  For $\tilde m > \tilde m_{\rm cr}$, the lower--most
branch of the curve corresponds to the stable phase, and the
upper--most branch of the curve corresponds to a metastable phase.
At $m=m_{\rm cr}$ we have a first order phase transition.  It
should be noted that the intermediate branch of the curve corresponds
to an unstable phase.

\begin{figure}[ht] %
   \centering
   \includegraphics[width=9cm]{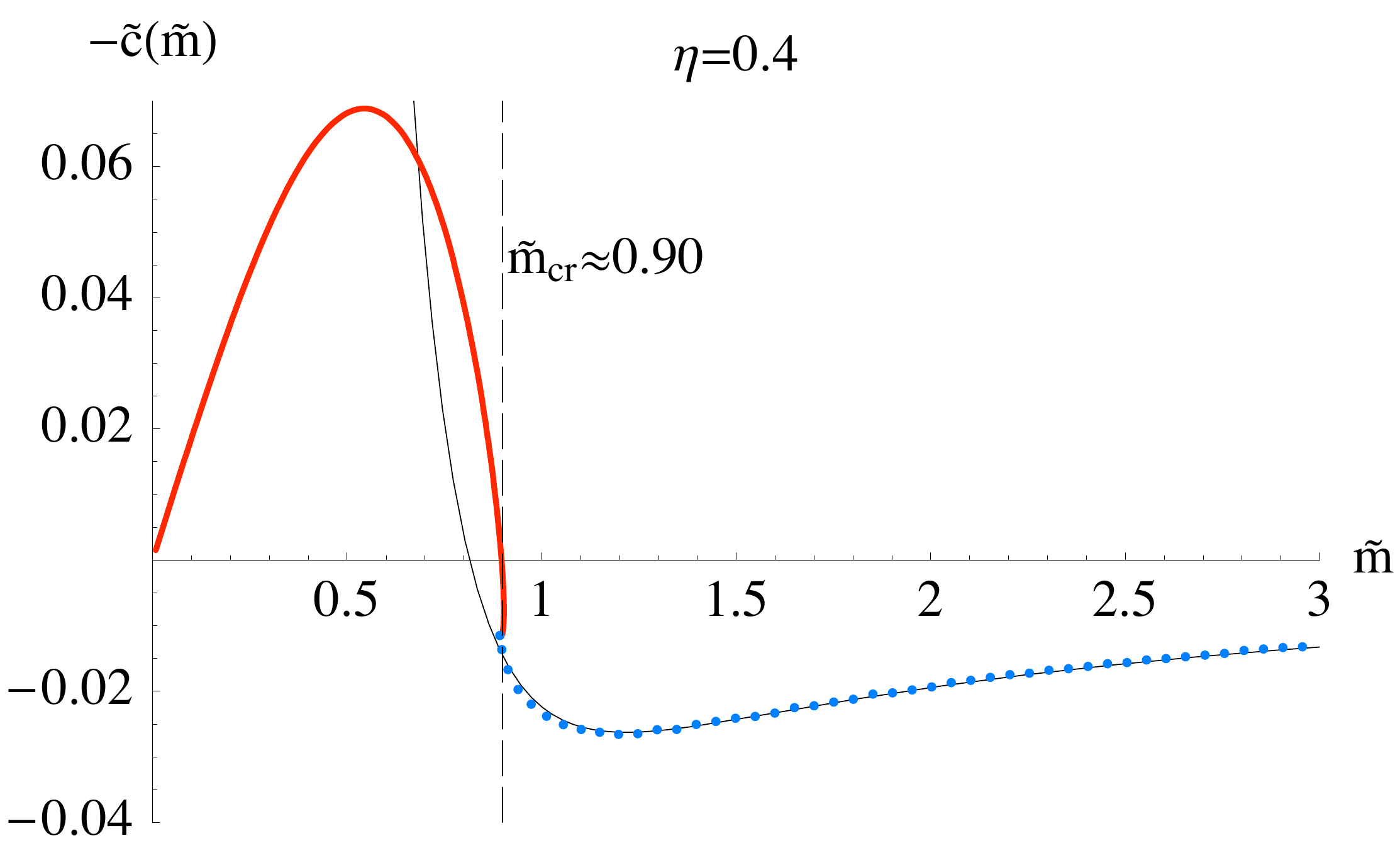} 
   \caption{\small The effect of the weak magnetic field is to decrease the values of $\tilde m_{\rm cr}$ and the condensate. Equation (\ref{weakcond}) is still a good approximation for $\tilde m>\tilde m_{\rm cr}$.}
   \label{fig:3}
\end{figure}

\begin{figure}[ht] %
   \centering
   \includegraphics[width=8cm]{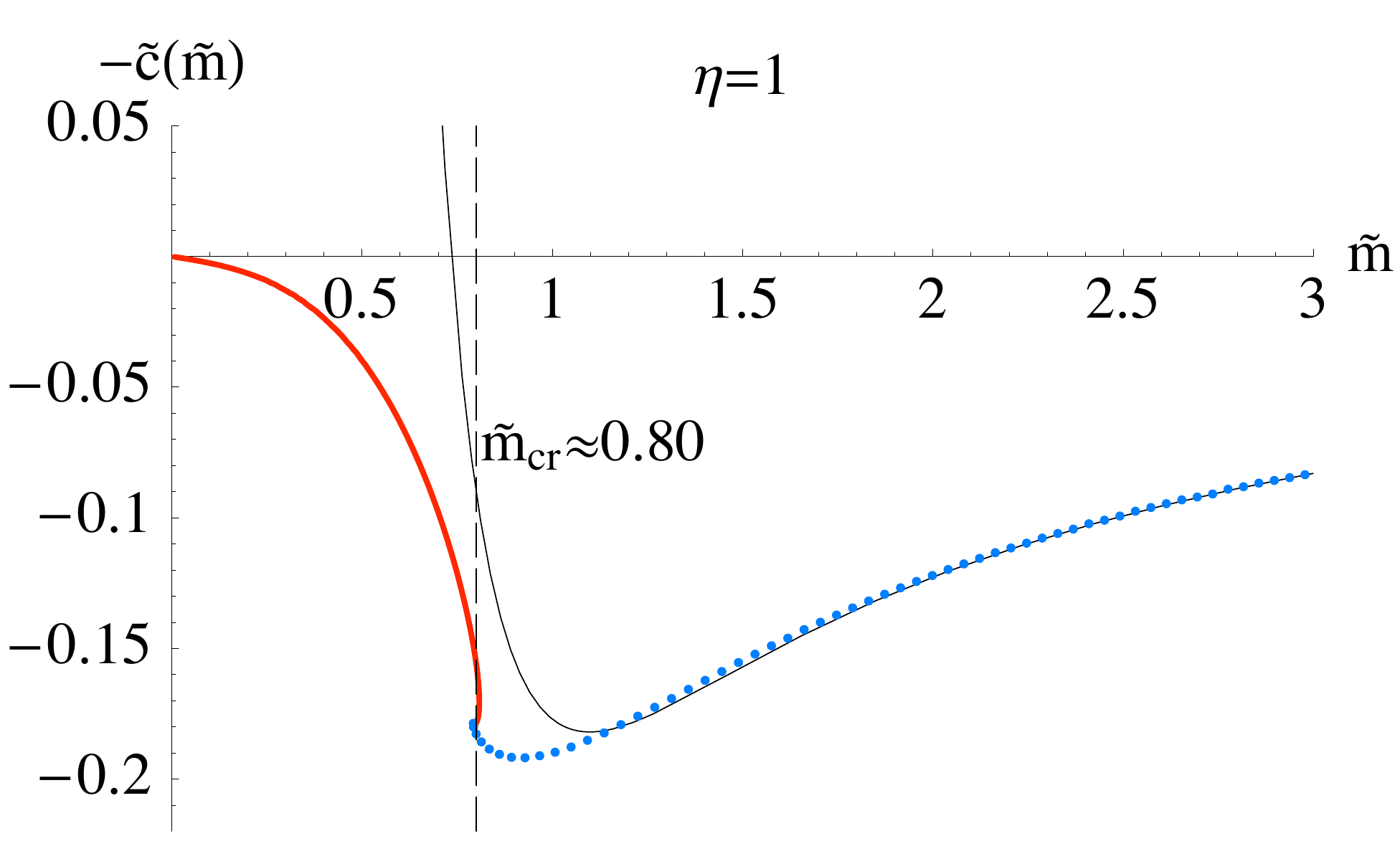} 
      \includegraphics[width=8cm]{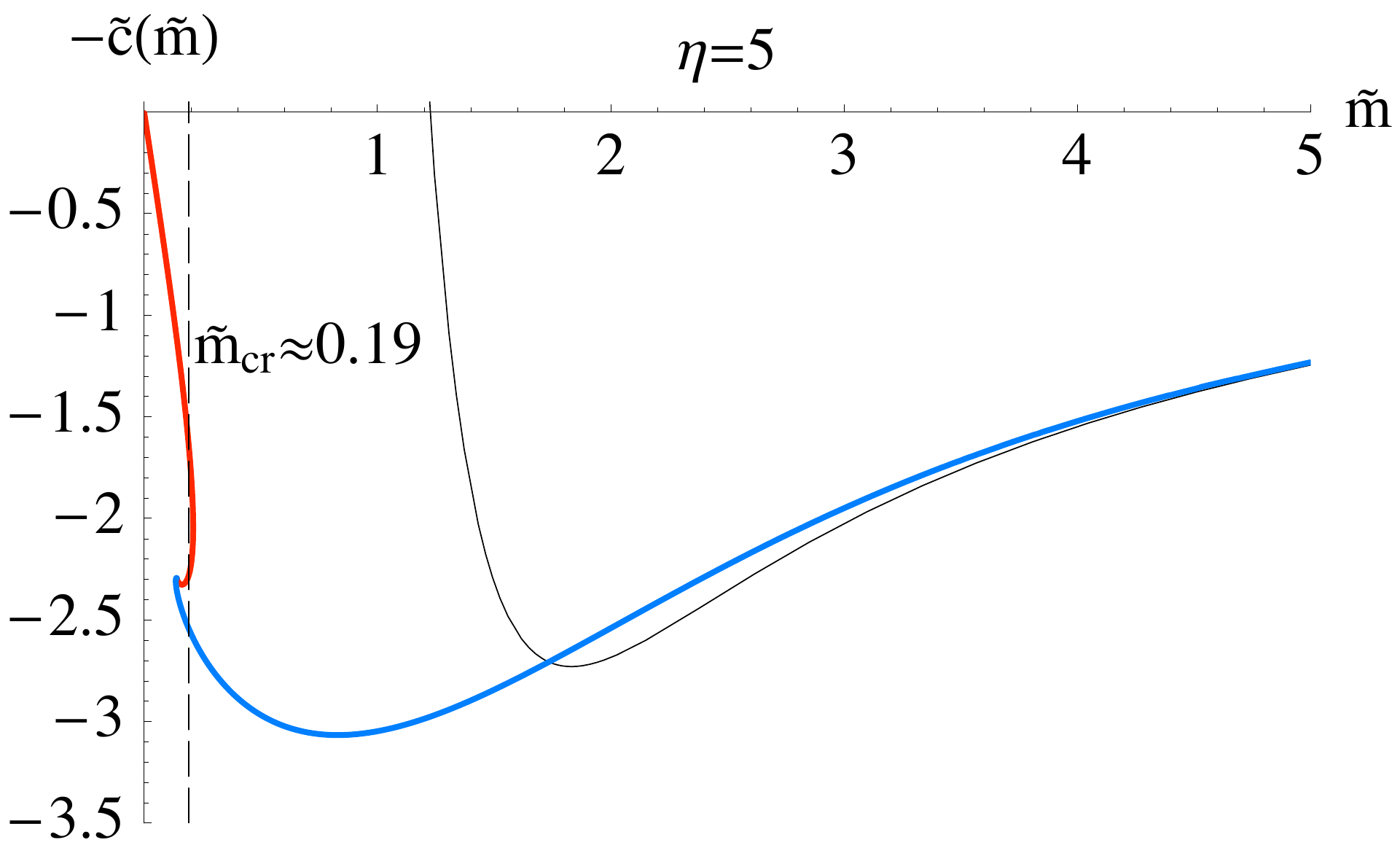} 
   \caption{\small For strong magnetic  field the condensate is negative. The value of $\tilde m_{\rm cr}$ continues to drop as we increase $\eta$.}
   \label{fig:4}
   \end{figure}

\begin{figure}[ht] 
   \centering
   \includegraphics[width=9cm]{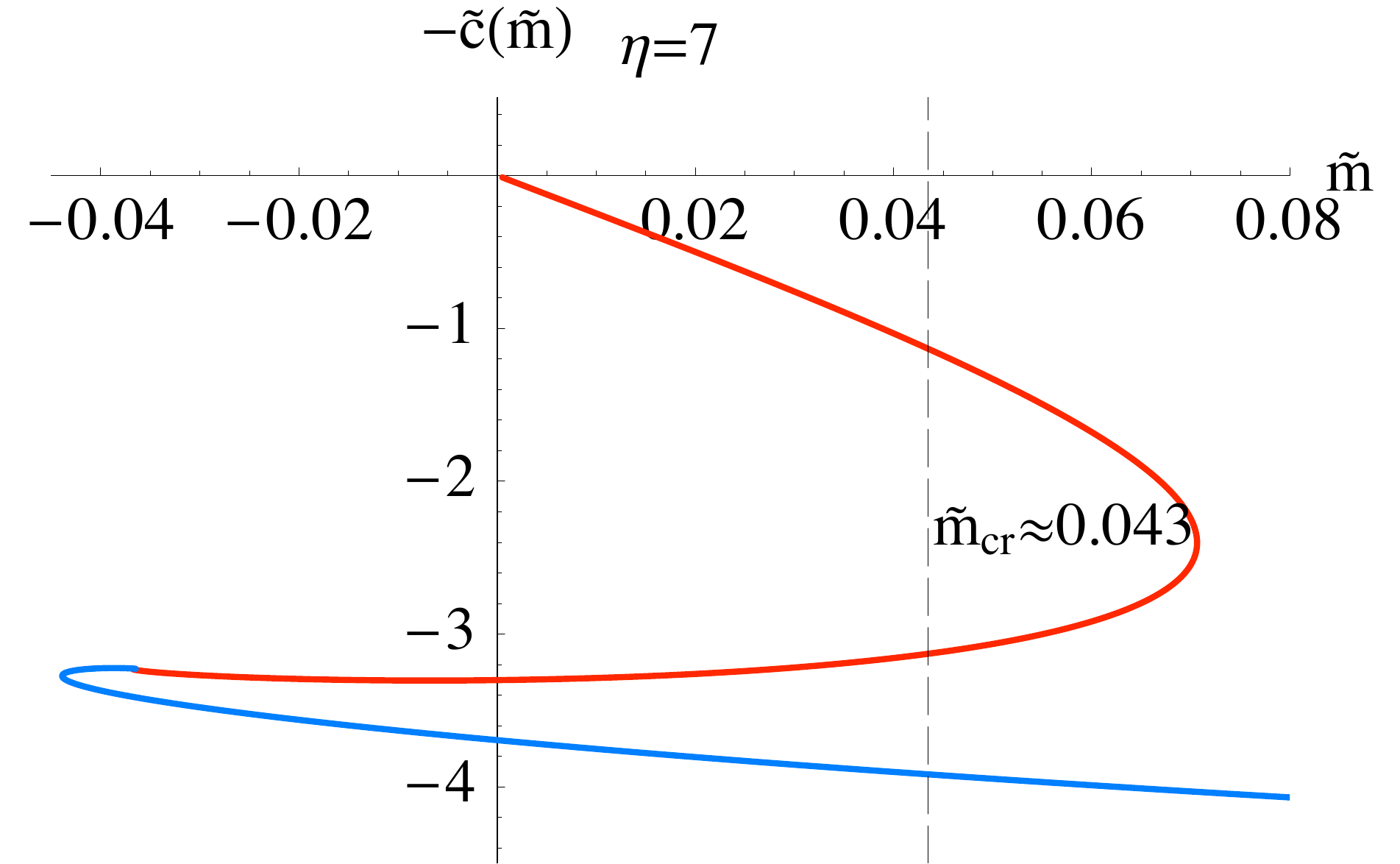} 
   \caption{\small For sufficiently high values of $\eta$ there are states with negative $\tilde m$, which are considered non-physical. However the equal area law is still valid as long as $m_{\rm cr}>0$.}
   \label{fig:5}
\end{figure}

\begin{figure}[ht] 
   \centering
   \includegraphics[width=10cm]{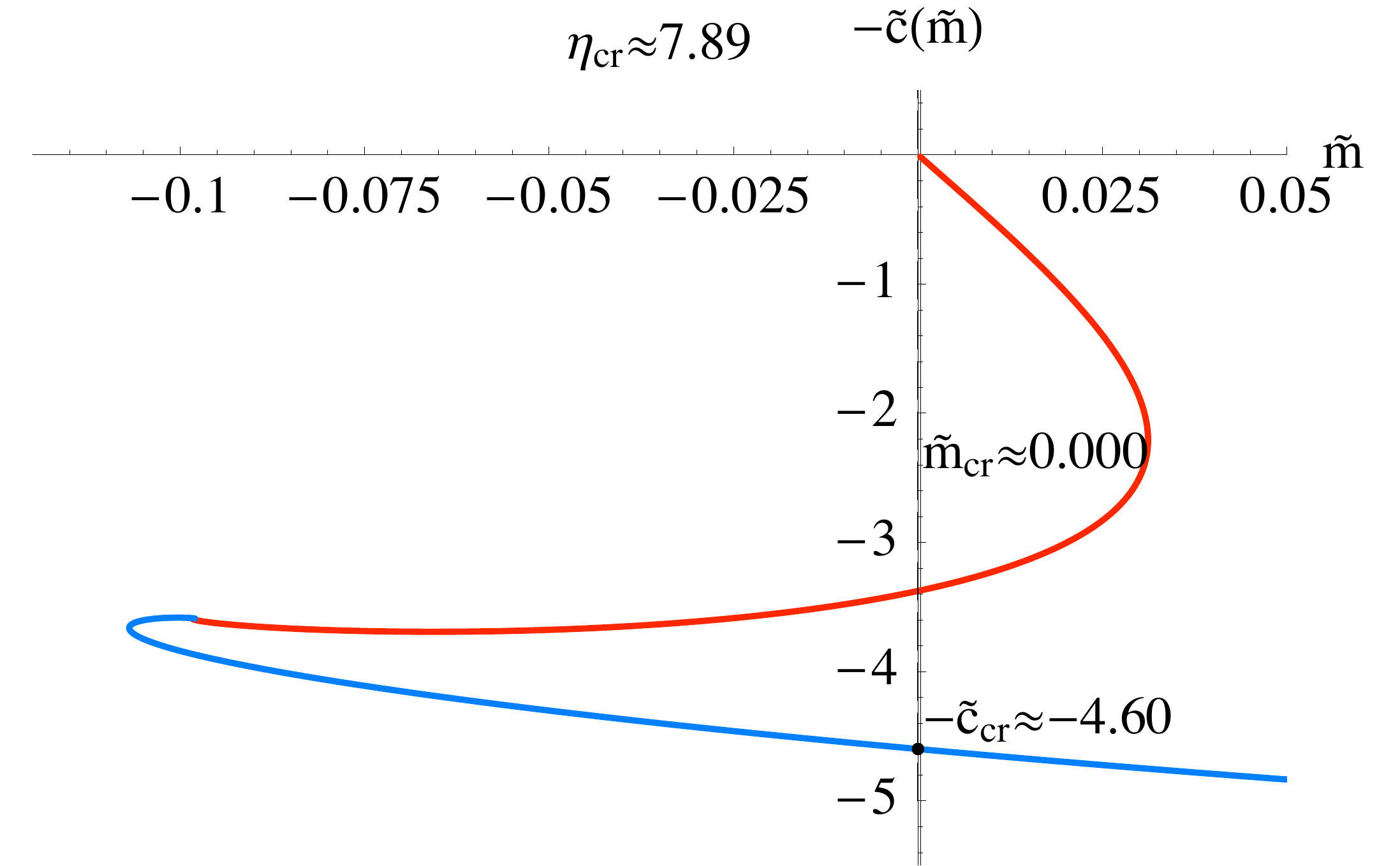} 
   \caption{\small For $\eta=\eta_{\rm cr}$ the critical parameter $m_{\rm cr}$ vanishes. There are two $\tilde m=0$ states with equal energies, one of them has non-vanishing condensate $-\tilde c_{\rm cr}\approx -4.60$ and therefore spontaneously breaks the chiral symmetry. }
   \label{fig:6}
\end{figure}

\begin{figure}[ht]
   \centering
   \includegraphics[width=10cm]{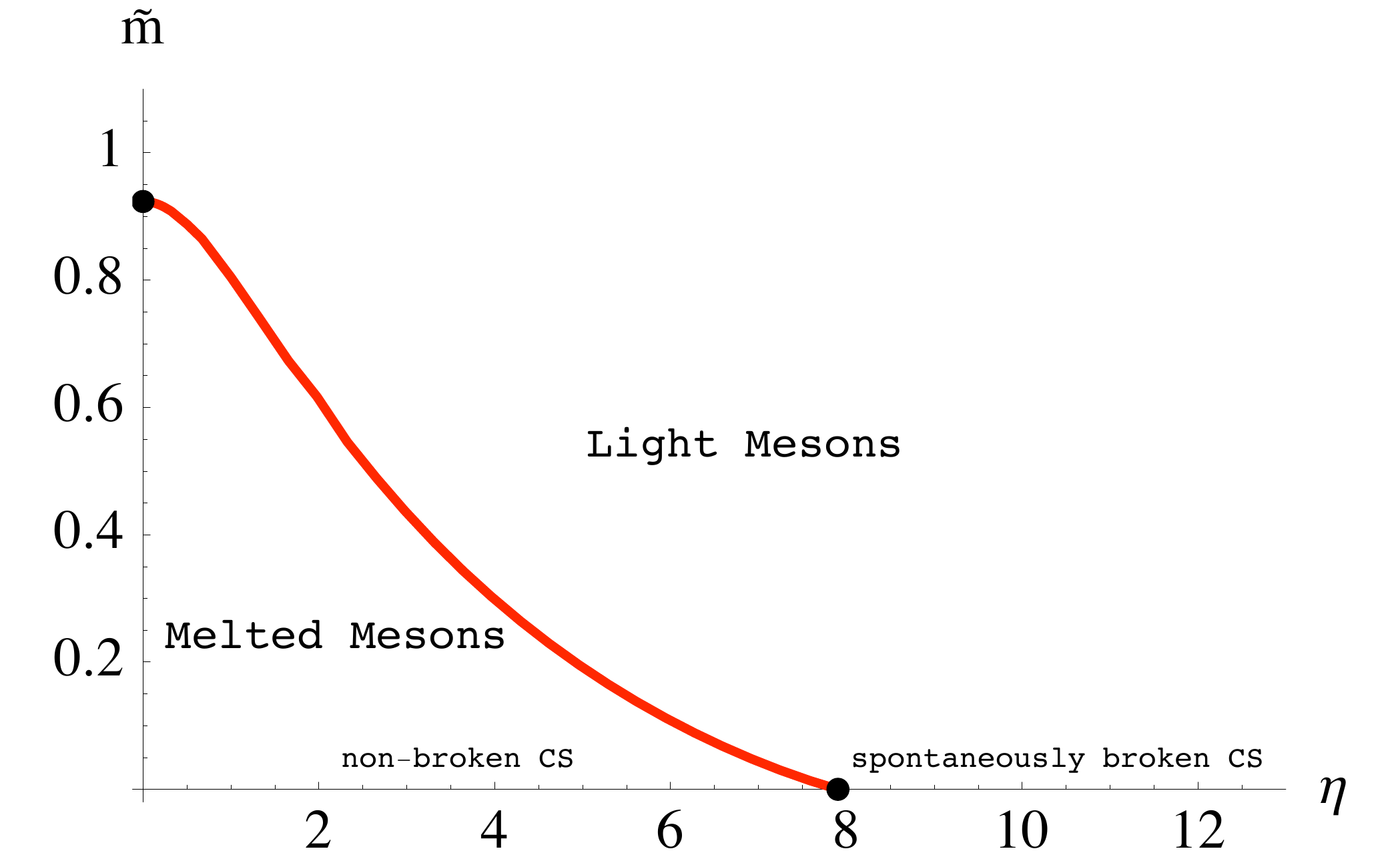} 
   \caption{\small The curve separates the two phases corresponding to discrete meson spectrum (light mesons) and continuous meson spectrum (melted mesons).}
   \label{fig:phase diagram}
\end{figure}

Now, let us turn on a weak magnetic field.  As one can see from figure
\ref{fig:3}, the effect of the magnetic field is to decrease the
magnitude of $\tilde m_{\rm cr}$.  In addition, the condensate now becomes
negative for sufficiently large $\tilde m$ and approaches zero from
below as $\tilde m\to\infty$. It is also interesting that equation
\reef{weakcond} is still a good approximation for $\tilde m>\tilde
m_{\rm cr}$.
For sufficiently strong magnetic field, the condensate has only
negative values and the critical value of $\tilde m$ continues to
decrease, as is presented in figure \ref{fig:4}.
   If we further increase the magnitude of the magnetic field, some
   states start having negative values of $\tilde m$, as shown in
   figure \ref{fig:5}. The negative values of $\tilde m$ do not mean
   that we have  negative bare quark masses; rather, it implies that
   the D7--brane embeddings have crossed $L = 0$ at least once. It was
   argued in ref.~\cite{Babington:2003vm} that such embeddings are not
   consistent with a holographic gauge theory interpretation and are
   therefore to be considered unphysical.  We will adopt this
   interpretation here, therefore taking as physical only the $\tilde
   m>0$ branch of the $-\tilde c$ vs $\tilde m$ plots.  However, the
   prescription for determining the value of $m_{\rm cr}$ continues to be
   valid, as long as the obtained value of $\tilde m_{\rm cr}$ is
   positive.  Therefore, we will continue to use it in order to
   determine the value of $\eta\equiv\eta_{\rm cr}$ for which $\tilde
   m_{\rm cr}=0$.
As one can see in figure \ref{fig:6}, the value of $\eta_{\rm cr}$ that we
obtain is $\eta_{\rm cr}\approx 7.89$. Note also that, for this value of
$\eta$, the Minkowski $\tilde m=0$ embedding has a non--zero fermionic
condensate $\tilde c_{\rm cr}$, and hence the chiral symmetry is
spontaneously broken. For $\eta>\eta_{\rm cr}$, the stable solutions are
purely Minkowski embeddings, and the first order phase transition
disappears; therefore, we have only one class of solutions (the blue
curve) that exhibit spontaneous chiral symmetry breaking at zero bare
quark mass.  Some black hole embeddings remain meta--stable, but
eventually all black hole embeddings become unstable for large enough
$\eta$.  This is confirmed by our study of the meson spectrum, which
we present in later sections of the paper.
The above results can be summarized in a single two dimensional phase
diagram, which we present in figure \ref{fig:phase diagram}.
The curve separates the two phases corresponding to a discrete meson
spectrum (light mesons) and a continuous meson spectrum (melted
mesons) respectively. The crossing of the  curve is associated with
the first order phase transition corresponding to the melting of the
mesons. If we cross the  curve along the vertical axis, we have the
phase transition described in refs.~\cite{Babington:2003vm,
  Mateos:2006nu, Albash:2006ew}. Crossing the curve along the
horizontal axis corresponds to a transition from unbroken to
spontaneously broken chiral symmetry\cite{Filev:2007gb}, meaning the
parameter $\tilde c$ jumps from zero to $\tilde c_{\rm cr}\approx 4.60$,
resulting in non--zero fermionic condensate of the ground state.
It is interesting to explore the dependence of the fermionic
condensate at zero bare quark mass on the magnetic field. From
dimensional analysis it follows that:
\begin{equation}
c_{\rm cr}=b^3\tilde c_{\rm cr}(\eta)=\frac{\tilde c_{\rm cr}(\eta)}{\eta^{3/2}}R^3H^{3/2} \ .
\label{ccrit}
\end{equation}
\begin{figure}[ht] 
   \centering
   \includegraphics[width=11cm]{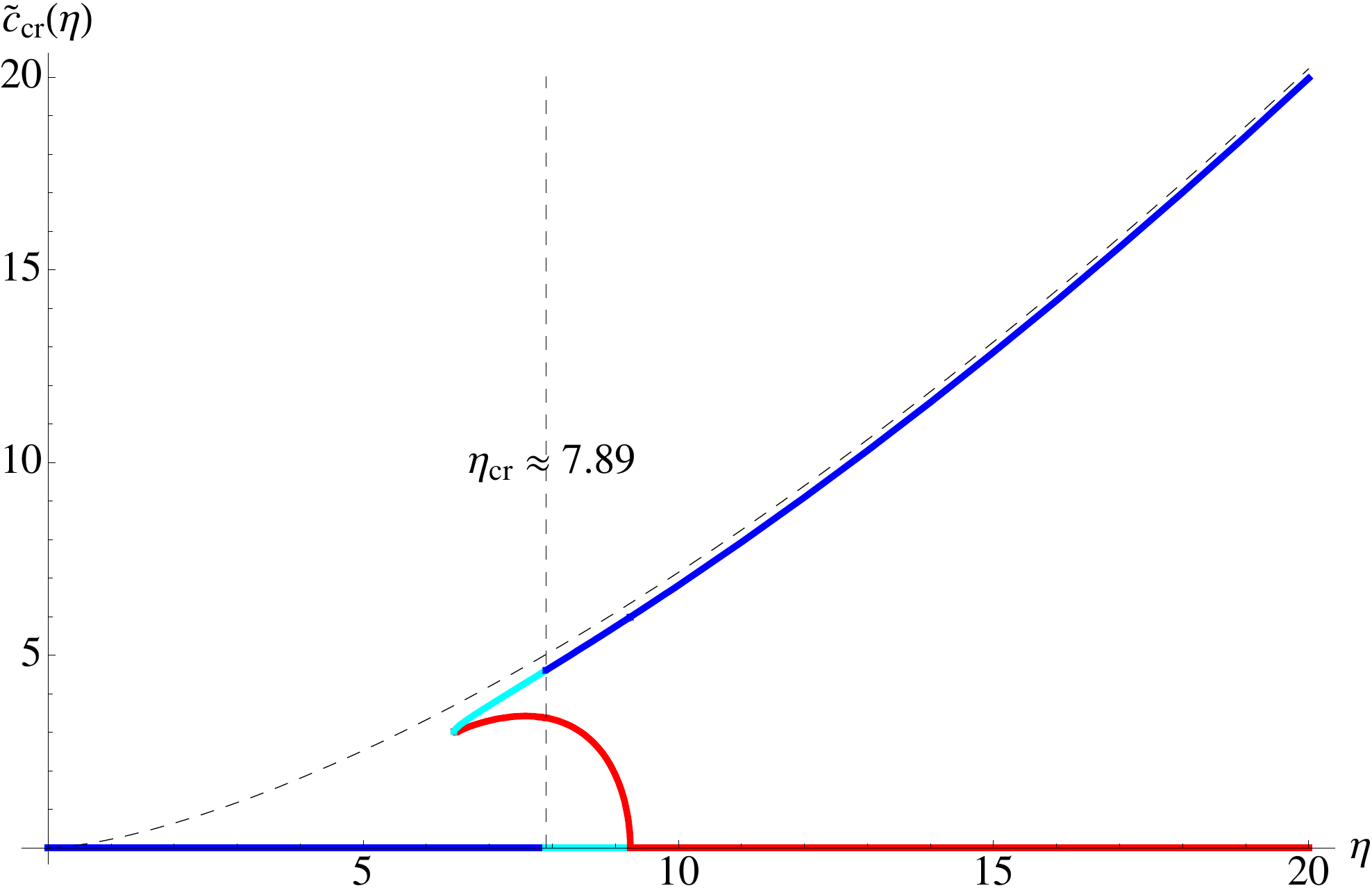} 
   \caption{\small The solid curves are the numerically extracted dependence $\tilde c_{\rm cr}(\eta)$, while the dashed curve represents the expected large $\eta$ behavior $\tilde c_{\rm cr}(\eta)\approx 0.226\eta^{3/2}$.  The solid  curve segments at the bottom left and to the upper right (blue) are the stable states. The straight segment and the arc that joins it (lower right, red) and red are the unstable states. The rest (cyan) are meta--stable states.}
   \label{fig:ccrit}
\end{figure}
In the $T\to 0$ limit, we should recover the result from
ref.~\cite{Filev:2007gb}: $c_{\rm cr}\approx0.226R^3H^{3/2}$, which
implies that $\tilde c_{\rm cr}(\eta)\approx 0.226\eta^{3/2}$ for $\eta\gg
1$.  The plot of the numerically extracted dependence $\tilde
c_{\rm cr}(\eta)$ is presented in figure \ref{fig:ccrit}; for
$\eta>\eta_{\rm cr}$, $\tilde c_{\rm cr}(\eta)$ very fast approaches the curve
$0.226\eta^{3/2}$. This suggests that the value of the chiral symmetry
breaking parameter $c_{\rm cr}$ depends mainly on the magnitude of the
magnetic field $H$, and only weakly on the temperature $T$.
%
\section{Thermodynamics}

Having understood the phase structure of the system, we now turn to
the extraction of various of its important thermodynamic quantities.
\subsection{The Free Energy}
%
Looking at our system from a thermodynamic point of view, we must
specify the potential characterizing our ensemble. We are fixing the
temperature and the magnetic field, and hence the appropriate
thermodynamic potential density is:
\begin{equation}
dF=-SdT-\mu dH\ ,
\end{equation}
where $\mu$ is the magnetization density and $S$ is the entropy
density of the system.  Following ref.~\cite{Mateos:2007vn}, we relate
the on--shell D7--brane action to the potential density~$F$ via:
\begin{equation}
F= 2 \pi^2N_f T_{D7}I_{D7}\ ,
\label{TDpot}
\end{equation}
where (here, $N_f=1$):
\begin{eqnarray}
&&I_{D7}=b^4\int\limits_{\tilde\rho_{\rm min}}^{\tilde\rho_{\rm max}}d\tilde\rho\tilde\rho^3\left(1-\frac{1}{16\tilde r^8}\right)\left(1+\frac{16\eta^2\tilde r^4}{(4\tilde r^4+1)^2}\right)^{\frac12}\sqrt{1+\tilde L'^2}+I_{\rm bound};
\label{action}\\
&&\eta=\frac{R^2}{b^2}H;~~~\tilde r=r/b;~~~\tilde\rho=\rho/b; ~~~\tilde L=L/b;~~~r^2=\rho^2+L^2 .\nonumber
\end{eqnarray}
In principle, on the right hand side of equation \reef{TDpot}, there
should be terms proportional to $-H^2/2$, which subtract the energy of
the magnetic field alone; however, as we comment below, the
regularization of $I_{D7}$ is determined up to a boundary term of the
form $\mathrm{const}\times H^2$.  Therefore, we can omit this term in
  the definition of $F$.
  The boundary action $I_{\rm bound}$ contains counterterms
  designed\cite{Balasubramanian:1999re} to cancel the divergent terms
  coming from the integral in equation \reef{action} in the limit of
  $\rho_{\rm max}\to\infty$.  A crucial observation is that the finite
  temperature does not introduce new divergences, and we have the
  usual quartic divergence from the spatial volume of the
  asymptotically AdS$_5$ spacetime~\cite{Karch:2005ms}. The presence
  of the non--zero external magnetic field introduces a new
  logarithmic divergence, which can be cancelled by introducing the
  following counterterm:
\begin{equation}
- \frac{R^4}{2}\log\left(\frac{\rho_{\rm max}}{R}\right) \int d^4x\sqrt{-\gamma}\frac{1}{2!}B_{\mu\nu}B^{\mu\nu} \ ,
\end{equation}
where $\gamma$ is the metric of the 4--dimensional surface at
$\rho=\rho_{\rm max}$. Note that in our case:
\begin{equation}
\frac{1}{2!}\sqrt{-\gamma}\, B_{\mu\nu}B^{\mu\nu}=H^2\ ,
\label{counterterm}
\end{equation}
which gives us the freedom to add finite terms of the form
$\mathrm{const}\times H^2$ at no cost to the regularized action.  This
makes the computation of some physical quantities scheme dependent.
We will discuss this further in subsequent sections.
The final form of $I_{\rm bound}$ in equation \reef{action} is:
\begin{equation}
I_{\rm bound}=-\frac{1}{4}\rho_{\rm max}^4-\frac{1}{2}R^4H^2\log{\frac{\rho_{\rm max}}{R}} \ .
\end{equation}
It is instructive to evaluate the integral in equation \reef{action}
for the $L\equiv0$ embedding at zero temperature.  Going back to
dimensionful coordinates we obtain:
\begin{equation}
\int\limits_{0}^{\rho_{\rm max}}d\rho\rho^3\sqrt{1+\frac{R^4H^2}{\rho^4}}=\frac{1}{4}\rho_{\rm max}^4+\frac{1}{2}R^4H^2\log{\frac{\rho_{\rm max}}{R}}+\frac{R^4H^2}{8}(1+\log{4}-\log{H^2})+O(\rho_{\rm max}^{-3}) \ .
\end{equation} 
The first two terms are removed by the counter terms from $I_{\rm bound}$,
and we are left with:
\begin{equation}
F(b=0,m=0,H)= 2 \pi^2 N_f T_{D7}\frac{R^4H^2}{8}(1+\log{4}-\log{H^2}) \ .
\label{freeenergy0}
\end{equation}
This result can be used to evaluate the magnetization density of the
Yang--Mills plasma at zero temperature and zero bare quark mass. Let
us proceed by writing down a more general expression for the free
energy of the system.  After adding the regulating terms from
$I_{\rm bound}$, we obtain that our free energy is a function of $m, b,
H$:
\begin{equation}
F(b,m,H)= 2 \pi^2 N_f T_{D7}b^4\tilde I_{D7}(\tilde m,\eta^2)+F(0,0,H) \ ,
\label{freeenergy_gen}
\end{equation}
where $\tilde I_{D7}(\tilde m,\eta)$ is defined {\it via}:
\begin {eqnarray}
  \tilde I_{D7} &=&\int\limits_{\tilde\rho_{\rm min}}^{\tilde\rho_{\rm max}}d\tilde\rho\left(\tilde\rho^3\left(1-\frac{1}{16\tilde r^8}\right)\left(1+\frac{16\eta^2\tilde r^4}{(4\tilde r^4+1)^2}\right)^{\frac12}\sqrt{1+\tilde L'^2}-\tilde\rho^3\right)-\tilde\rho_{\rm min}^4/4\label{tildeaction}\\
  &&-\frac{1}{2}\eta^2\log\tilde\rho_{\rm max}-\frac{1}{8}\eta^2(1+\log4-\log\eta^2);~~~\tilde
  r^2=\tilde\rho^2+\tilde L(\tilde\rho)^2\nonumber \ .
\end{eqnarray}
In order to verify the consistency of our analysis with our numerical
results, we derive an analytic expression for the free energy that is
valid for $\tilde m\gg \sqrt{\eta}$. To do
this we use that for large $\tilde m$, the condensate $\tilde c$ is
given by equation \reef{weakcond}, which we repeat here:
\begin{equation}
\tilde c(\tilde m,\eta^2)=\frac{\eta^2}{4\tilde m}-\frac{1+4\eta^2+8\eta^4}{96\tilde m^5}+O(1/\tilde m^7) \ ,
\end{equation}
as well as the relation ${\partial\tilde I_{D7}}/{\partial\tilde
  m}=-2\tilde c$.  We then have:
%
\begin{equation}
\tilde I_{D7}=-2\int\limits^{\tilde m}\tilde c(\tilde m,\eta)d\tilde m+\xi(\eta)=\xi(\eta)-\frac{1}{2}\eta^2\log\tilde m-\frac{1+4\eta^2+8\eta^4}{192\tilde m^4}+O(1/\tilde m^6) \ ,
\label{tildeaction1}
\end{equation}
where the function $\xi(\eta)$ can be obtained by evaluating the
expression for $\tilde I_{D7}$ from equation \reef{tildeaction} in the
approximation $\tilde L\approx \tilde m$. Note that this suggests
ignoring the term $\tilde L'^2$, which is of order $\tilde c^2$.
Since the leading behavior of $\tilde c^2$ at large $\tilde m$ is
$1/\tilde m^2$, this means that the results obtained by setting
$\tilde L'^2=0$ can be trusted to the order of $1/\tilde m$, and
therefore we can deduce the function $\xi(\eta)$, corresponding to the
zeroth order term. Another observation from earlier in this paper is
that the leading behavior of the condensate is dominated by the
magnetic field and therefore we can further simplify equation
\reef{tildeaction}:
\begin{eqnarray}
\tilde I_{D7}&\backsimeq &\lim_{\tilde\rho_{\rm max}\to\infty}\int\limits_0^{\tilde\rho_{\rm max}}d\tilde\rho\rho^3\left(\sqrt{1+\frac{\eta^2}{(\tilde\rho^2+\tilde m^2)^2}}-1\right)-\frac{1}{2}\eta^2\log\tilde\rho_{\rm max}-\frac{1}{8}\eta^2\left(1-\log\frac{\eta^2}{4}\right)\nonumber\\
&=&-\frac{\eta^2}{2}\log\tilde m-\frac{\eta^2}{8}(3-\log\frac{\eta^2}{4})+O(1/\tilde m^3)\ .
\end{eqnarray}
Comparing to equation \reef{tildeaction1}, we obtain:
\begin{equation}
\xi(\eta)=-\frac{\eta^2}{8}(3-\log\frac{\eta^2}{4}) \ ,
\end{equation}
%
and our final expression for $\tilde I_{D7}$, valid for $\tilde m\gg \sqrt{\eta}$:
\begin{equation}
\tilde I_{D7}=-\frac{\eta^2}{8}\left(3-\log\frac{\eta^2}{4}\right)-\frac{1}{2}\eta^2\log\tilde m-\frac{1+4\eta^2+8\eta^4}{192\tilde m^4}+O(1/\tilde m^6) \ .
\label{tildeaction2}
\end{equation}
%
\subsection{The Entropy}
\begin{figure}[ht]
 \centering
  \includegraphics[width=11cm]{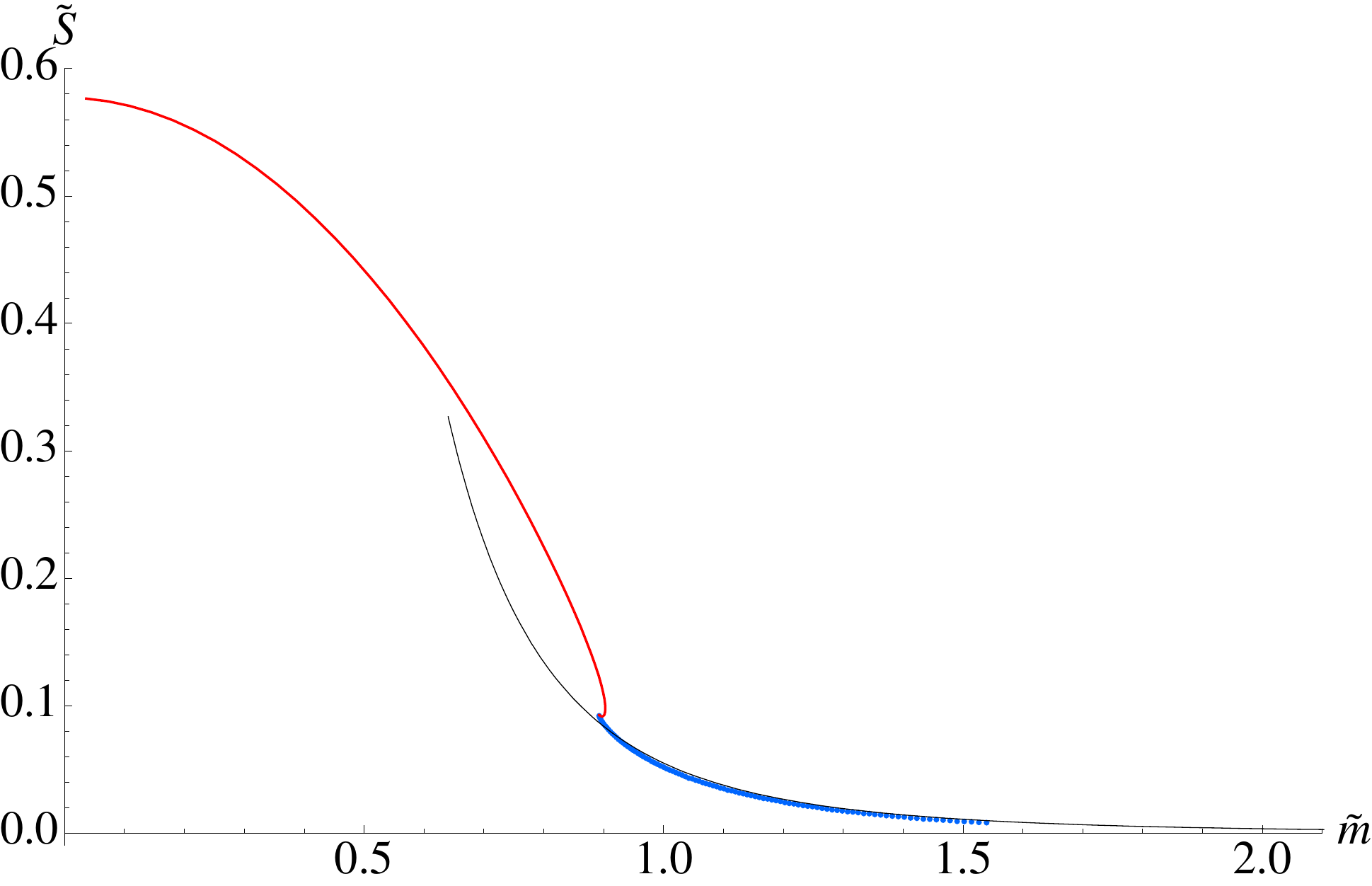} 
   \caption{\small \small A  plot of $\tilde S$ versus $\tilde m$ for $\eta=0.4$ . The  thin (sharply descending and extending to the right) black curve corresponds to that large mass result of equation~(\ref{large m}).}
   \label{fig:entropyh=.4}
 \end{figure}

Our next goal is to calculate the entropy density of the system. Using
our expressions for the free energy we can write:
\begin{eqnarray}
S&=&-\left(\frac{\partial F}{\partial T}\right)_{H}=-\pi R^2 \frac{\partial F}{\partial b}=-2 \pi^3 R^2 N_f T_{D7}b^3\left(4\tilde I_{D7}+b\frac{\partial\tilde I_{D7}}{\partial\tilde m}\frac{\partial\tilde m}{\partial b}+b\frac{\partial\tilde I_{D7}}{\partial\eta^2}\frac{\partial\eta^2}{\partial b}\right) \nonumber \\
&=&-2 \pi^3 R^2 N_f T_{D7}b^3\left(4\tilde I_{D7}+2\tilde c\tilde m-4\frac{\partial\tilde I_{D7}}{\partial\eta^2}\eta^2\right)=2 \pi^3 R^2 N_f T_{D7}b^3\tilde S(\tilde m,\eta^2)\ . \label{entropy}
\end{eqnarray}
It is useful to calculate the entropy density at zero bare quark mass and zero fermionic condensate. To do this, we need to calculate the free energy density by evaluating the integral in equation \reef{tildeaction} for $\tilde L\equiv0$. The expression that we get for $\tilde I_{D7}(0,\eta^2)$ is:
\begin{equation}
\tilde I_{D7}(0,\eta^2)=\frac {1} {8}\left (1-2 \sqrt {1 + \eta^2}-\eta^2\log\frac{(1+ \sqrt {1 + \eta^2})^2}{\eta^2} \right)\ .
\end{equation} 
The corresponding expression for the entropy density is:
\begin{equation}
S|_{m=0}=2 \pi^6 R^8 N_f T_{D7}T^3\left(-\frac{1}{2}+\sqrt{1+\frac{\pi^4 H^2}{R^4 T^4}}\right)\ .
\end{equation}
One can see that the entropy density is positive and goes to zero as
$T\to 0$. Our next goal is to solve for the entropy density at finite
$\tilde m$ for fixed $\eta$. To do so, we have to integrate
numerically equation \reef{entropy} and generate a plot of $\tilde S$
versus $\tilde m$. However, for $\tilde m\gg \sqrt{\eta}$ we can
derive an analytic expression for the entropy.  After substituting the
expression from equation \reef{tildeaction2} for $\tilde I_{D7}$ into
equation \reef{entropy} we obtain:
 \begin{equation}
 \tilde S(\tilde m,\eta^2)=\frac{1+2\eta^2}{24\tilde m^4}+\dots \ ,
 \label{large m}
 \end{equation}
 or if we go back to dimensionful parameters:
 \begin{equation}
 S(b,m,H)=2 \pi^3R^2 N_f T_{D7}b^3\left(\frac{b^4+2R^4H^2}{24m^4}\right)\ .
 \end{equation}
 One can see that if we send $T\to 0$, while keeping $\eta$ fixed we
 get the $T^7$ behavior described in ref.~\cite{Mateos:2007vn}, and
 therefore the (approximate; $N_f/N\ll1$) conformal behaviour is
 restored in this limit.  In figure \ref{fig:entropyh=.4}, we present
 a plot of $\tilde S$ versus $\tilde m$ for $\eta=0.4$ . The solid
 smooth black curve corresponds to equation \reef{large m}.  For this
 $\tilde S$ is positive and always a decreasing function of $\tilde
 m$.  Hence, the entropy density at fixed bare quark mass $m=\tilde m
 b$, given by $S=2 \pi^3 R^2 N_f T_{D7}m^3\tilde S/\tilde m^3$, is
 also a decreasing function of $\tilde m$ and therefore an increasing
 function of the temperature, except near the phase transition (the
 previously described crossover from black hole to Minkowski
 embeddings) where an unstable phase appears that is characterized by
 a negative heat capacity.
 %
 \subsection{The Magnetization}
\begin{figure}[ht] 
   \centering
   \includegraphics[width=10cm]{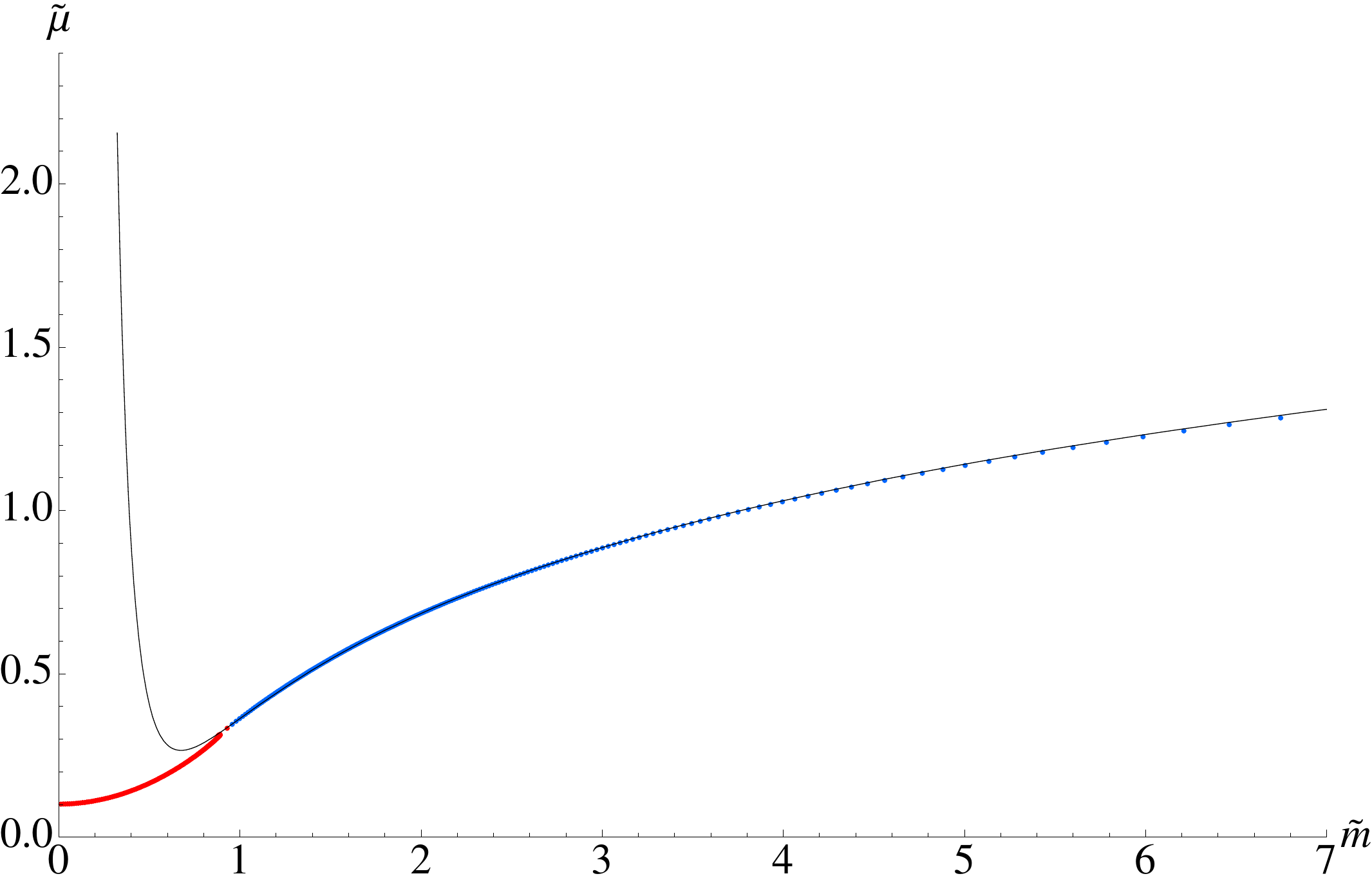} 
   \caption{\small \small A plot of the dimensionless relative magnetization $\tilde\mu$ versus $\tilde m$ for $\eta=0.5$. The thin black curve (starting with a steep descent) corresponds to the large mass  result of equation (\ref{magn-anal}).}
   \label{fig:magnh=0.5}
\end{figure}

 Let us consider equation \reef{freeenergy0} for the free energy
 density at zero temperature and zero bare quark mass. The
 corresponding magnetization density is given by:
\begin{equation}
\mu_0=-\left(\frac{\partial F}{\partial H}\right)_{T,m=0}=2 \pi^2 R^4 N_f T_{D7}\frac{H}{2}\log\frac{H}{2}.
\end{equation}
Note that this result is scheme dependent, because of the freedom to
add terms of the form $\mathrm{const}\times H^2$ to the boundary action that we
discussed earlier. However, the value of the relative magnetization is
given by:
\begin{equation}
\mu-\mu_0=-\left(\frac{\partial F}{\partial H}\right)_T-\mu_0=-2 \pi^2 R^2 N_f T_{D7}b^2 \left(\frac{\partial\tilde I_{D7}}{\partial\eta}\right)_{\tilde{m}}= 2 \pi^2 R^2 N_f T_{D7}b^2\tilde\mu \ ,
\label{magn}
\end{equation}
is scheme independent and is the quantity of interest in the section.
In equation \reef{magn}, we have defined
$\tilde\mu=-{\partial\tilde
    I_{D7}}/{\partial\eta}|_{\tilde{m}}$ as a dimensionless
parameter characterizing the relative magnetization.  Details of how
the derivative is taken are discussed in appendix \ref{appendix:eta}.
The expression for $\tilde\mu$ follows directly from equation
\reef{tildeaction}:
\begin{equation}
\tilde\mu=\lim_{\tilde\rho_{\rm max}\to\infty}-\int\limits_{\tilde\rho_{\rm min}}^{\tilde\rho_{\rm max}}d\tilde\rho\frac{\tilde\rho^3(4\tilde r^4-1)}{\tilde r^4\sqrt{(4\tilde r^4+1)^2+16\eta\tilde r^4}}+\eta\log\tilde\rho_{\rm max}-\frac{\eta}{2}\log\frac{\eta}{2}.
\label{tildemagnetization}
\end{equation}
For the large $\tilde m$ region we use the asymptotic expression for
$\tilde I_{D7}$ from equation \reef{tildeaction2} and obtain the
following analytic result for $\tilde\mu$:
\begin{equation}
 \tilde\mu=\frac{\eta}{2}-\frac{\eta}{2}\log\frac{\eta}{2}+\eta\log\tilde m+\frac{\eta(1+4\eta^2)}{24\tilde m^4}+O(1/\tilde m^6)\ .
\label{magn-anal}
\end{equation}
We evaluate the above integral numerically and generate a plot of
$\tilde\mu$ versus $\tilde m$.  A plot of the dimensionless relative
magnetization $\tilde\mu$ versus $\tilde m$ for $\eta=0.5$ is
presented in figure \ref{fig:magnh=0.5}. The black curve corresponding
to equation \reef{magn-anal} shows good agreement with the asymptotic
behavior at large $\tilde m$.  It is interesting to verify the
equilibrium condition ${\partial\tilde\mu}/{\partial T}>0$. Note
that since $\mu_0$ does not depend on the temperature, the value of
this derivative is a scheme independent quantity. From equations
\reef{magn} and equation \reef{magn-anal}, one can obtain:
\begin{equation}
\frac{\partial\mu}{\partial T}=2 \pi^3 R^4 N_f T_{D7}b\left(2\tilde\mu-\frac{\partial\tilde\mu}{\partial\tilde m}\tilde m-2\frac{\partial\tilde\mu}{\partial\eta}\eta\right)=2 \pi^3 R^6 N_f T_{D7}\frac{Hb^3}{6m^4}>0\ ,
\end{equation}
which is valid for large $m$ and weak magnetic field $H$. Note that
the magnetization seems to increase with the temperature. Presumably
this means that the temperature increases the ``ionization'' of the
Yang--Mills plasma of mesons even before the phase transition occurs.
\subsection{The Speed of Sound}
\begin{figure}[!ht]
\begin{center}
\subfigure[] {\includegraphics[angle=0,
width=0.45\textwidth]{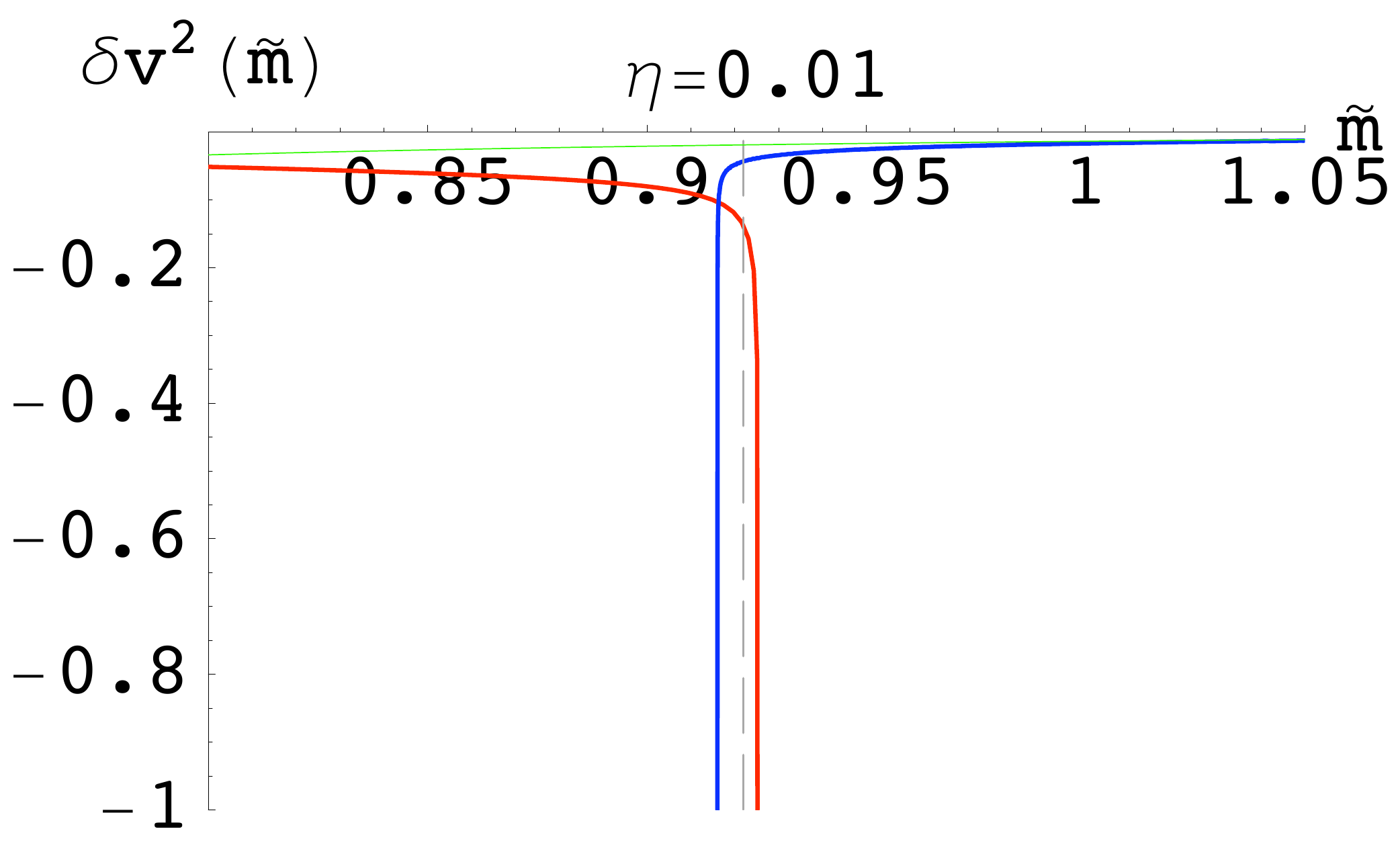} \label{fig: h001deltav}}
\subfigure[] {\includegraphics[angle=0,
width=0.45\textwidth]{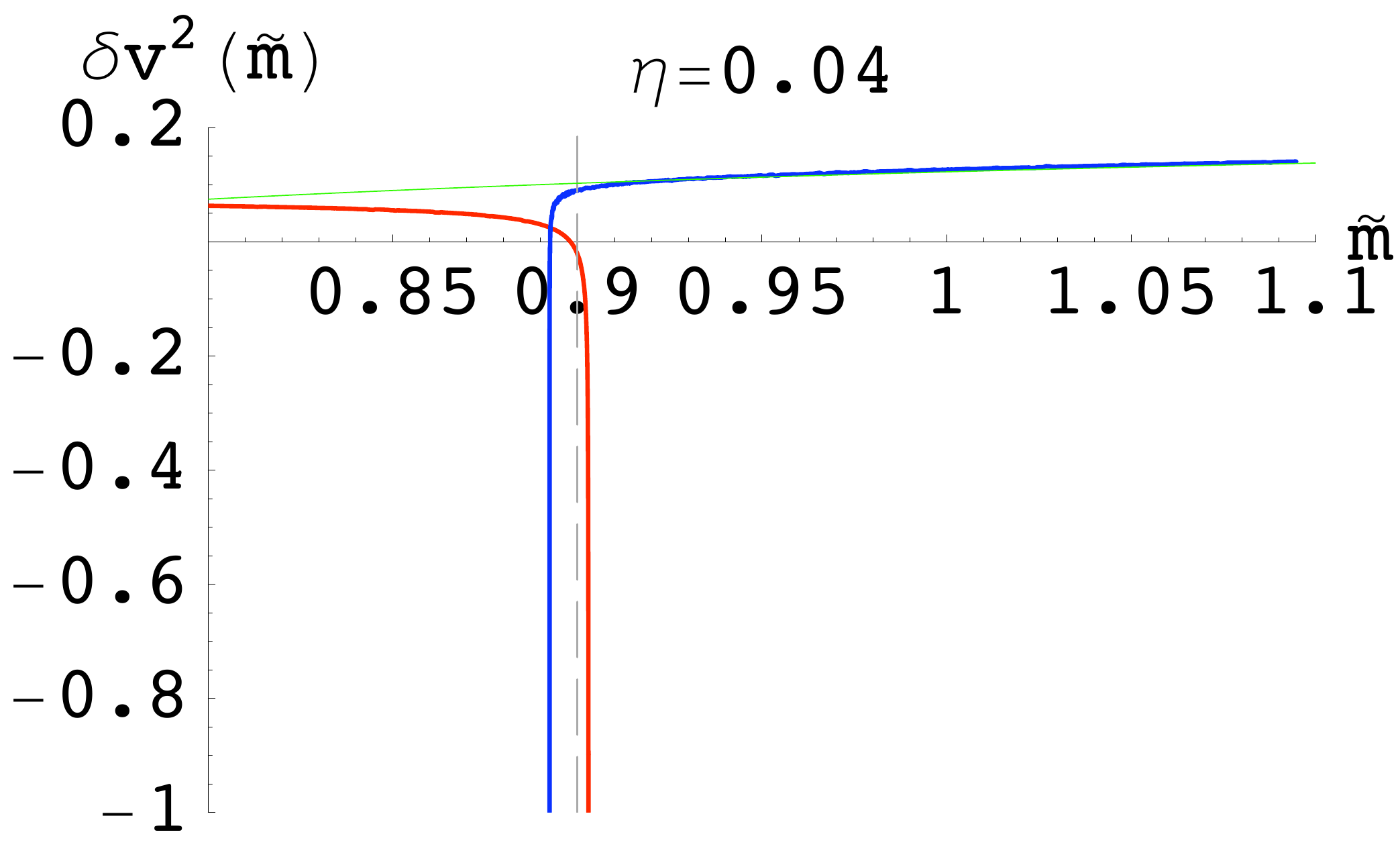} \label{fig: h004deltav}}
\subfigure[] {\includegraphics[angle=0,
width=0.45\textwidth]{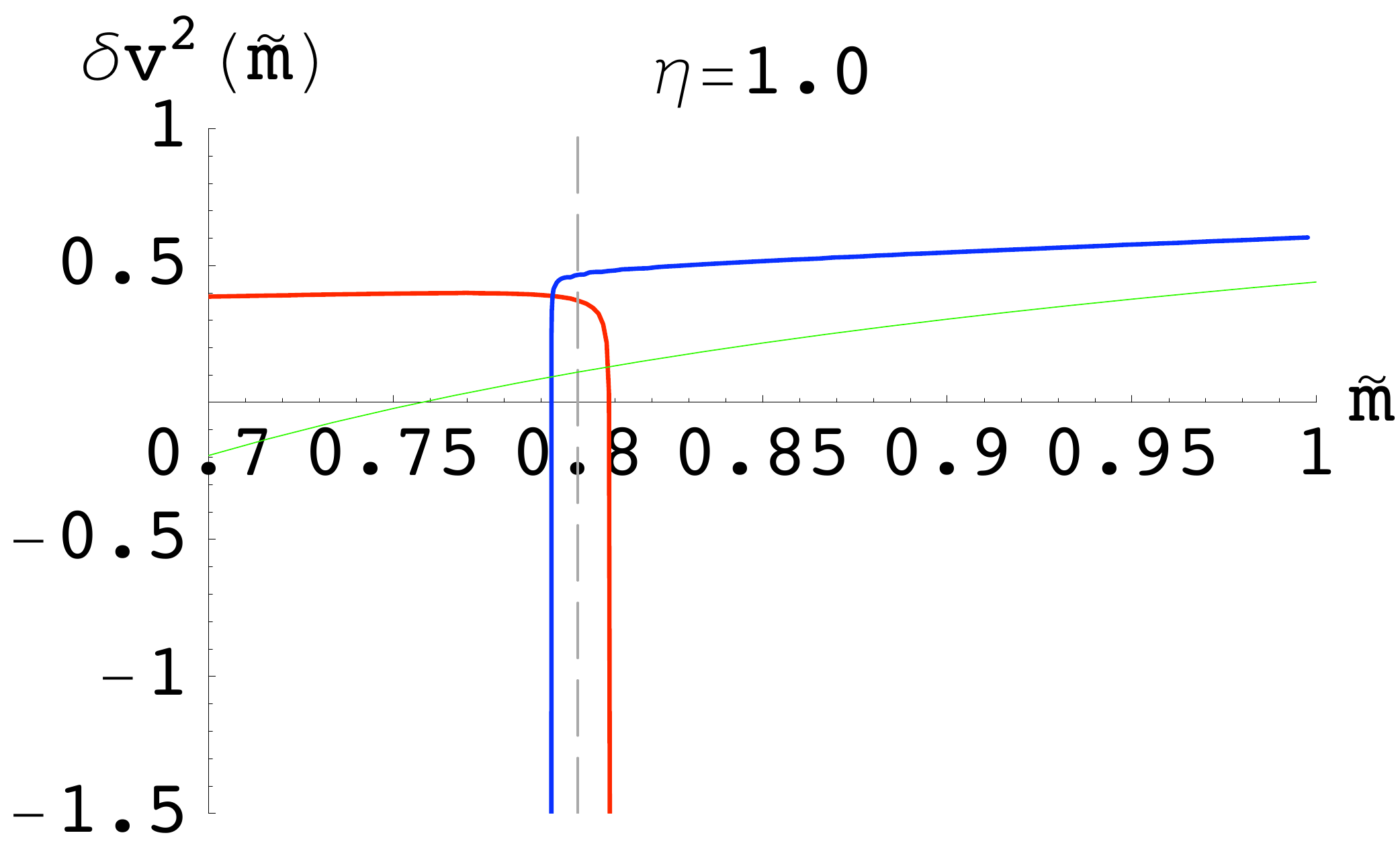} \label{fig: h01deltav}}
\subfigure[] {\includegraphics[angle=0,
width=0.45\textwidth]{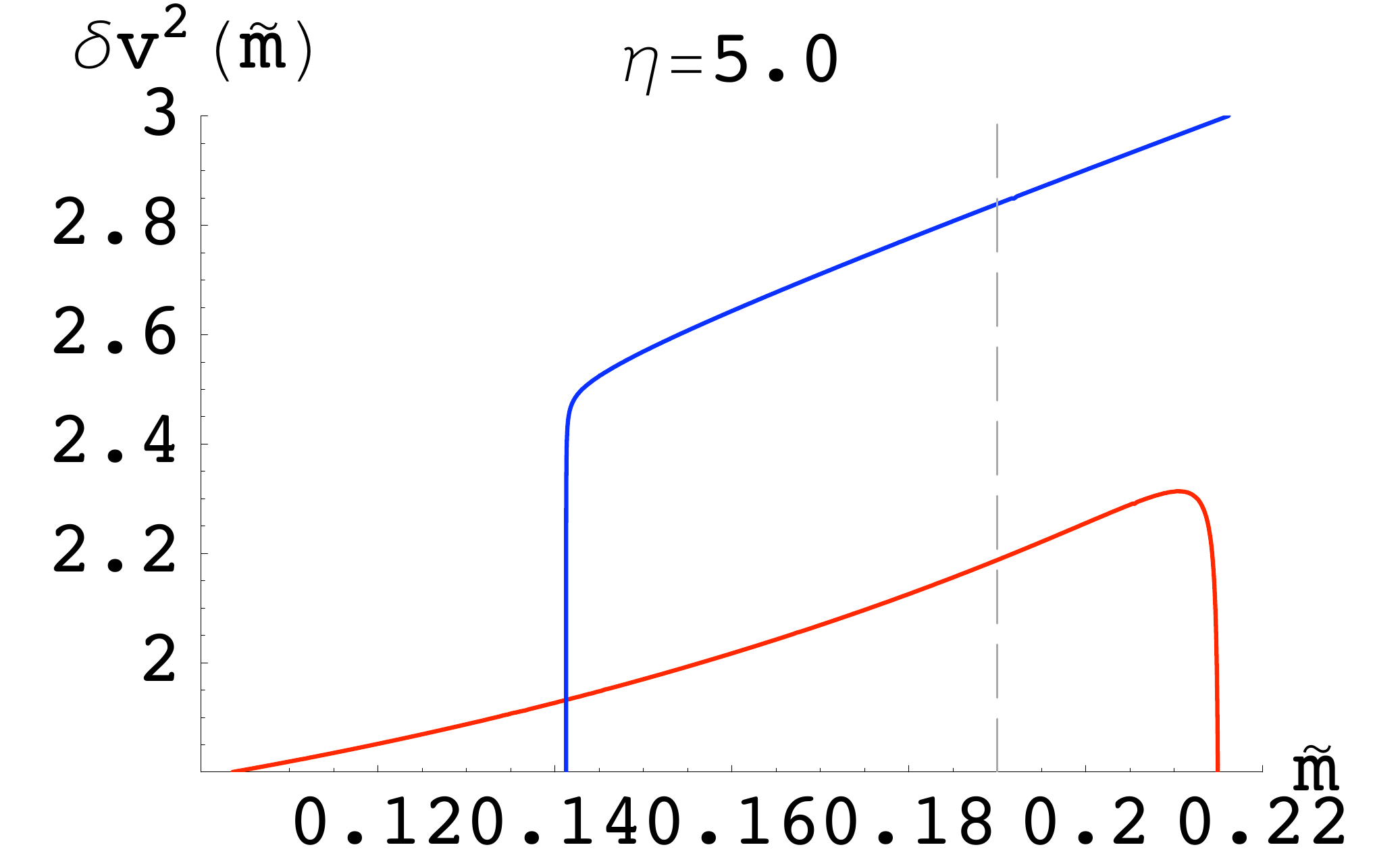} \label{fig: h05deltav}}
\caption{\small The deviation of the speed of sound from the conformal value in units of $(\nu\lambda)\pi^2/3$ in the presence of fundamental matter and an external magnetic field. The  curves coming in from the left (red) correspond to black hole embeddings, and the  curves coming in from the right (blue) correspond to Minkowski embeddings. The vertical dashed line represents the phase transition point, and the flatter  dashed curves (green) correspond to the approximate analytic expression given in \reef{eqt: deviationapp}. We do not include the  curve of the analytic result in \ref{fig: h05deltav} since the approximate formula is not valid for high magnetic fields.}
\label{fig: sdev}
\end{center}
\end{figure}

It is interesting to investigate the effect of the magnetic field on
the speed of sound in the Yang--Mills plasma. Following
ref.~\cite{Mateos:2007vn}, we use the following definition to
thermodynamically determine the speed:
\begin{eqnarray}\label{eqt: sounddef}
v^2&=&\frac{S}{c_V}=\frac{S_{D3}+S_{D7}}{c_{V3}+c_{V7}} \ ,
\end{eqnarray}
where $c_V$ is the density of the heat capacity at constant volume. To
compute the contribution coming from the fundamental flavors in the
presence of an external magnetic field, we work perturbatively in small
$N_f/N_c$.  First, let us recall the adjoint contribution to
entropy and specific heat \cite{Mateos:2007vn}:
\begin{eqnarray}
S_{D3}&=&-\frac{\pi^2}{2}N^2T^3\ , \quad c_{V3}=3S_{D3}\ .
\end{eqnarray}
To proceed, let us rewrite the entropy density of the fundamental
flavours in the following form:
\begin{eqnarray}\label{eqt: sfrel} 
S_{D7} &=& -\frac{4F}{T}\left(1+\frac{2\mathcal{\widetilde N}\tilde{m}\tilde{c}(\pi T)^4}{4F}\right)+\frac{4}{T}\left(F_0+\mathcal{\widetilde N}(\pi T)^4\eta^2\frac{\partial\tilde{I}_{D7}}{\partial\eta^2}\right),
\end{eqnarray}
where $\mathcal{\widetilde N}=2\pi^2 N_f T_{D7}$.  The term $4 \left(F_0-F
\right)/T$ is simply the contribution from the conformal theory; the
deviation from it is related to the conformal symmetry being broken by
introducing the fundamental flavors and the external magnetic field.
This breaking is manifest by non--vanishing $\tilde{c}$ and $\eta$ in
equation \reef{eqt: sfrel} respectively. Recalling the relation
between the energy density and the free energy density $E=F+TS$, and,
using equation \reef{eqt: sfrel}, we find that:
\begin{eqnarray}
c_{V7} &=& \left(\frac{\partial E}{\partial T}\right)_V \nonumber \\
&=&3S_{D7}-2\mathcal{\widetilde N}\pi^4\frac{\partial}{\partial T}(T^4\tilde{m}\tilde{c})+4\mathcal{\widetilde N}\pi^4\eta^2\frac{\partial}{\partial T}\left(T^4\frac{\partial\tilde{I}_{D7}}{\partial\eta^2}\right).
\end{eqnarray}
Using the definition in equation~\reef{eqt: sounddef}, together with
the above results and expanding up to first order in $\nu = N_f/N_c$, we
get:
\begin{equation}\label{eqt: soundsp}
v^2 \approx \frac{1}{3}\left[1+\frac{\lambda N_f}{N_c}\frac{\pi^2}{6}\left(\tilde{m}\tilde{c}-\frac{1}{3}\tilde{m}^2\frac{\partial\tilde{c}}{\partial\tilde{m}}\right)-\frac{\lambda N_f}{N_c}\frac{\pi^2\eta^2}{3}\left(\frac{4}{3}\frac{\partial\tilde{I}_{D7}}{\partial\eta^2}+\frac{1}{3}\frac{\partial}{\partial\eta^2}(2\tilde{m}\tilde{c})\right)\right]+O(\nu^2).
\end{equation}
The second and third term in equation \reef{eqt: soundsp} represent
the deviation from the conformal value of $1/3$ by the presence of the
fundamental flavors and the external magnetic field.
For convenience let us define $\delta v^2=v^2-{1}/{3}$. It is possible
to obtain an analytic expression for $\delta v^2$ in the limit of
large bare quark mass and small magnetic field ($\tilde m\gg \sqrt{\eta}$). Using our previous analytic expressions, we get:
\begin{eqnarray}\label{eqt: deviationapp}
\delta v^2\approx \lambda\frac{N_f}{N}\frac{\pi^2}{3}\left(\frac{2}{3}\eta^2\log\tilde{m}-\frac{1}{6}\eta^2\log\left(\frac{\eta^2}{4}\right)-\frac{1-24\tilde{m}^4\eta^2+8\eta^4}{72\tilde{m}^4}\right)+O\left(\frac{1}{\tilde{m}^5}\ .\right)
\end{eqnarray}
It is important to note that equation \reef{eqt: deviationapp} is
valid only up to first order in $\nu$.  To proceed beyond the large
bare quark mass and small magnetic field limit, we study numerically
the velocity deviation, which is summarised in figure \ref{fig: sdev}.
We observe from figure \ref{fig: h001deltav} that, for small magnetic
field, the deviation is similar to the zero magnetic field case;
$\delta v^2$ approaches zero (corresponding to restoration of the
conformal symmetry) from below in both the $T\to 0$ and $T\to\infty$
limits. However in presence of large magnetic field (see figure
\ref{fig: h05deltav}), we see that $\delta v^2>0$, and the conformal
value is never attained.
\section{Meson Spectrum}
In this section, we calculate the meson spectrum of the gauge theory.
The mesons we are considering are formed from quark--antiquark pairs,
so the relevant objects to consider are 7--7 strings.  In our
supergravity description, these strings are described by fluctuations
(to second order in $\alpha'$) of the probe branes' action about the
classical embeddings we found in the previous
sections\cite{Kruczenski:2003be}.  Studying the meson spectrum serves
two purposes.  First, tachyons in the meson spectrum from fluctuations
of the classical embeddings indicate the instability of the embedding.
Second, a massless meson satisfying a Gell-Mann-Oakes-Renner (GMOR)
relation will confirm that spontaneous chiral symmetry breaking has
occurred.  As a reminder, in ref.~\cite{Kruczenski:2003be}, the exact
meson spectrum for the AdS$_5 \times S^5$ background was found to be
given by:
\begin{eqnarray}
M(n,\ell) &=& \frac{2 m}{R^2} \sqrt{(n+\ell + 1)(n + \ell + 2)} \ ,
\end{eqnarray}
where $\ell$ labels the order of the spherical harmonic expansion, and $n$ is a positive integer that represents the order of the mode.
The relevant pieces of the action to second order in $\alpha'$ are:
\begin{eqnarray} \label{eqt:fluctuation}
S/N_f&=& -  T_{D7} \int d^8 \xi \sqrt{g_{ab} + B_{a b}+ 2 \pi \alpha' F_{a b}} + \left(2\pi \alpha'\right) \mu_7 \int_{\mathcal{M}_8} F_{(2)} \wedge B_{(2)} \wedge P\left[\tilde{C}_{(4)} \right]  \nonumber \\
&&  +\left(2\pi \alpha' \right)^2 \mu_7 \frac{1}{2} \int_{\mathcal{M}_8} F_{(2)} \wedge F_{(2)} \wedge P\left[C_{(4)}\right] \ , \\
C_{(4)} &=& \frac{1}{g_s} \frac{u^4}{R^4} dt \wedge dx^1 \wedge dx^2 \wedge dx^3  \ ,\end{eqnarray}
\begin{equation}
\tilde{C}_{(4)} = -\frac{R^4}{g_s}  \left(1- \cos^4\theta \right) \sin\psi \cos\psi\ d \psi \wedge d \phi_2 \wedge d \phi_3 \wedge d\phi _1\ ,
\end{equation}
where $P\left[C_{(4)}\right]$ is the pull--back of the 4--form
potential sourced by the stack of $N_c$ D3--branes,
$P\left[\tilde{C}_{(4)}\right]$ is the pull--back of the 4--form
magnetic dual to $C_{(4)}$, and $F_{(2)}$ is the Maxwell 2--form on
the D7--brane worldvolume.  At this point, we resort to a different
set of coordinates than we have been using.  Instead of using the
coordinates $(\rho, L)$ introduced in
equation~\reef{eqt:changeofcoordinates}, we return to the coordinates
$(z=1/u^2, \theta)$ because the analysis is simpler.
We consider fluctuations of the form:
\begin{eqnarray}
\theta &=& \theta_0(z) + 2 \pi \alpha' \chi(\xi^a) \ , \label{eqt:ansatz1} \\
\phi_1 &=& 2 \pi \alpha' \Phi(\xi^a) \ , \label{eqt:ansatz2}
\end{eqnarray}
where the indices $a, b = 0 \dots 7$ run along the worldvolume of the
D7--brane.  $\theta_0(z)$ corresponds to the classical embedding from
the classical equations of motion.  Plugging the ansatz in
equations~\reef{eqt:ansatz1} and~\reef{eqt:ansatz2} into the action
and expanding to second order in $\left(2 \pi \alpha' \right)$, we get
as second order terms in the lagrangian:
\begin{eqnarray}
-\mathcal{L}_{\chi^2} &=& \frac{1}{2} \sqrt{-E} S^{a b} R^2 \partial_a \chi \partial_b \chi -\frac{1}{2} \sqrt{-E} R^4 \left(\theta_0' \right)^2 E^{z z} S^{a b} \partial_a \chi \partial_b \chi \nonumber  \\
&& + \frac{1}{2} \chi^2 \left[ \partial_\theta^2 \sqrt{-E} - \partial_z \left(E^{zz} R^2 \theta_0' \partial_\theta \sqrt{-E} \right) \right] \nonumber \ , \\
-\mathcal{L}_{\Phi^2}&=&  \frac{1}{2} \sqrt{-E} S^{a b} R^2 \sin^2 \theta_0 \partial_a \Phi \partial_b \Phi \ , \nonumber \\
-\mathcal{L}_{F^2} &=& \frac{1}{4} \sqrt{-E} S^{a b} S^{c d} F_{b c} F_{a d}  \nonumber \ , \\ 
-\mathcal{L}_{F-\chi} &=& \chi F_{23} \left[ \partial_z \left(\sqrt{-E} R^2 \theta_0' E^{zz} J^{23} \right) - J^{23} \partial_\theta \sqrt{-E} \right] = \chi F_{2 3} f \nonumber \ , \\
\mathcal{L}_{F^2}^{\mathrm{WZ}} &=&  \frac{1}{8} \frac{1}{z^2 R^4} F_{m n} F_{o p} \epsilon^{m n o p} \nonumber \ , \\
\mathcal{L}_{F-\Phi}^{\mathrm{WZ}} &=& -\Phi F_{01} B_{23}  R^4 \sin\psi \cos\psi \partial_z \left( 1- \cos^4 \theta_0 \right) = - \Phi F_{0 1} B_{2 3} R^4 \sin \psi \cos \psi \partial_z K \ .
\end{eqnarray}
We have taken $E_{a b} = g^{(0)}_{a b} + B_{a b}$ to be the zeroth
order contribution from the DBI action.  In addition, we use that
$E^{a b} = S^{a b} + J^{a b}$, where $ S^{a b} = S^{ b a}$ and $J^{a
  b} = - J^{b a}$.  We use this notation for brevity.  The indices $m,
n, o, p = 4 \dots 9$ run in the transverse directions to the
D3--branes.  From these lagrangian terms, we derive the equation of
motion for $\chi$ to be:
\begin{eqnarray}
0&=& \partial_a \left(\sqrt{-E} S^{a b} R^2 \left(\frac{1+4b^4 z^4 \left(\theta_0'\right)^2}{1+4 z^2 \left(\theta'_0\right)^2} \right) \partial_b \chi \right) -\chi  \left[ \partial_\theta^2 \sqrt{-E} - \partial_z \left(E^{zz} R^2 \theta_0' \partial_\theta \sqrt{-E} \right) \right]  \nonumber \\
&& - F_{23} \left[ \partial_z \left(\sqrt{-E} R^2 \theta_0' E^{zz} J^{23} \right) - J^{23} \partial_\theta \sqrt{-E}  \right]  \ .
\end{eqnarray}
The equation of motion for $\Phi$ is given by:
\begin{eqnarray}
\partial_a \left( \sqrt{-E} S^{a b} R^2 \sin^2 \theta_0 \partial_b \Phi \right) - F_{01} B_{23}  R^4 \sin\psi \cos\psi \partial_z K &=& 0 \ .
\end{eqnarray}
The equation of motion for $A_b$ is given by: 
\begin{eqnarray}
 \partial_a \left( - \sqrt{-E} S^{a a'} S^{b b'} F_{a' b'} - \chi f  \left(\delta^a_2 \delta^b_3 - \delta^a_3 \delta^b_2 \right) + B_{23} \Phi \partial_z K \left(\delta^a_0 \delta^b_1 - \delta^a_1 \delta^b_0 \right) \right. && \nonumber  \\ 
 \left. + \frac{1}{2} \frac{1}{z^2 R^4} \epsilon^{ m n o p} \delta^a_m \delta^b_n F_{o p} \right) &=& 0 \ .
\end{eqnarray}
We are allowed to set $A_m = 0$ with the constraint (using that $S^{22} = S^{33}$):
\begin{eqnarray*}
S^{00} \partial_m \partial_0 A_0 + S^{11} \partial_m \partial_1 A_1 + S^{22} \partial_m \left(\partial_2 A_2 +  \partial_3 A_3\right) &=& 0  
\end{eqnarray*}
Therefore, we can consistently take $A_0 = \partial_1 A_1 = 0$, $\partial_2 A_2 = -\partial_3 A_3$.  With this particular choice, we have as equations of motion for the gauge field:
\begin{eqnarray*}
- \partial_0 \left(\sqrt{-E} S^{00} S^{11} \partial_0 A_1 \right) + \partial_z K B_{2 3} \partial_0 \Phi - \partial_z \left(\sqrt{-E} S^{zz} S^{11} \partial_z A_1 \right) &&\\
- \partial_{\tilde{m}} \left(\sqrt{-E} S^{\tilde{m} \tilde{n}} S^{11} \partial_{\tilde{n}} A_1 \right) &=& 0 \ ,\\
- \partial_0 \left(\sqrt{-E} S^{00} S^{22} \partial_0 A_2 \right) + f \partial_3 \chi - \partial_z \left(\sqrt{-E} S^{zz} S^{22} \partial_z A_2 \right) &&\\
- \partial_{\tilde{m}} \left(\sqrt{-E} S^{\tilde{m} \tilde{n}} S^{22} \partial_{\tilde{n}} A_2 \right) &=& 0 \ , \\
- \partial_0 \left(\sqrt{-E} S^{00} S^{33} \partial_0 A_3 \right) - f \partial_2 \chi - \partial_z \left(\sqrt{-E} S^{zz} S^{33} \partial_z A_3 \right) &&\\
- \partial_{\tilde{m}} \left(\sqrt{-E} S^{\tilde{m} \tilde{n}} S^{33} \partial_{\tilde{n}} A_3 \right) &=& 0 \ ,
\end{eqnarray*}
where the indices $\tilde{m}, \tilde{n}$ run over the $S^3$ that the D7--brane wraps.  If we assume that $\partial_i \chi = 0$, we find that the equations for $A_2$ and $A_3$ decouple from $\chi$.  Therefore, we can consistently take $F_{2 3} = 0$, or, in other words, $A_2 = A_3 = 0$.   This simplifies the equations of motion that we need to consider to:
\begin{eqnarray}
0 &=& \partial_a \left[\sqrt{-E} S^{a b} R^2 \left(\frac{1+4b^4 z^4 \left(\theta_0'\right)^2}{1+4 z^2 \left(\theta'_0\right)^2} \right) \partial_b \chi \right] -\chi  \left[ \partial_\theta^2 \sqrt{-E} \right. \nonumber \\
&& \left. - \partial_z \left(E^{zz} R^2 \theta_0' \partial_\theta \sqrt{-E} \right) \right]  \ , \label{eqt:eom_theta} \\
0 &=&- \partial_0 \left(\sqrt{-E} S^{00} S^{11} \partial_0 A_1 \right) + \partial_z K B_{2 3} \partial_0 \Phi - \partial_z \left(\sqrt{-E} S^{zz} S^{11} \partial_z A_1 \right)  \nonumber \\
&&- \partial_{\tilde{m}} \left(\sqrt{-E} S^{\tilde{m} \tilde{n}} S^{11} \partial_{\tilde{n}} A_1 \right)  \ , \label{eqt:eom_phi_A2} \\
0 &=& \partial_a \left( \sqrt{-E} S^{a b} R^2 \sin^2 \theta_0 \partial_b \Phi \right) - F_{01} B_{23}  R^4 \sin\psi \cos\psi \partial_z K  \ . \label{eqt:eom_phi_A1} 
\end{eqnarray}
In the proceeding sections, we will work out the solutions to these
equations numerically using a shooting method.  With an appropriate
choice of initial conditions at the event horizon, which we explain
below, we numerically solve these equations as an initial condition
problem in Mathematica.  Therefore, the D.E. solver routine ``shoots''
towards the boundary of the problem, and we extract the necessary data
at the boundary.
\subsection{The $\chi$ Meson Spectrum}
\begin{figure}[ht] 
\begin{center}
\includegraphics[width=10cm]{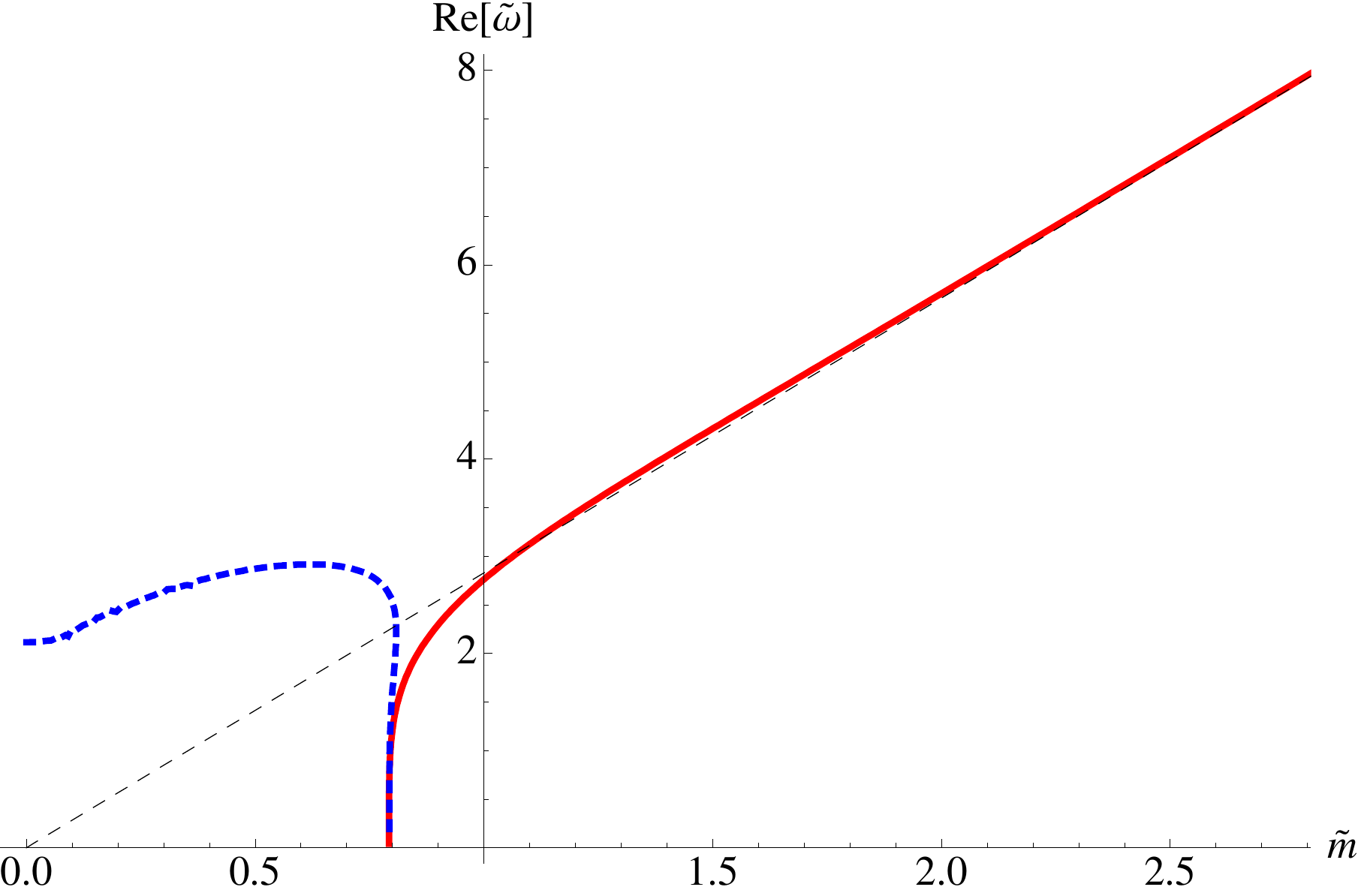}
\end{center}
\caption{\small The $\chi$ meson mass as a function of bare quark mass for $\eta = 1$.  The dashed (blue) curve corresponds to fluctuations about black hole embeddings.  The solid (red) line corresponds to fluctuations about Minkowski emeddings.  These modes have a purely real $\omega$.  The straight dashed (black) line corresponds to the pure AdS$_5 \times S^5$ solution.}  
\label{fig:fluctuations in theta eta1}
\end{figure}

\begin{figure}[ht] 
\begin{center}
  \includegraphics[width=10cm]{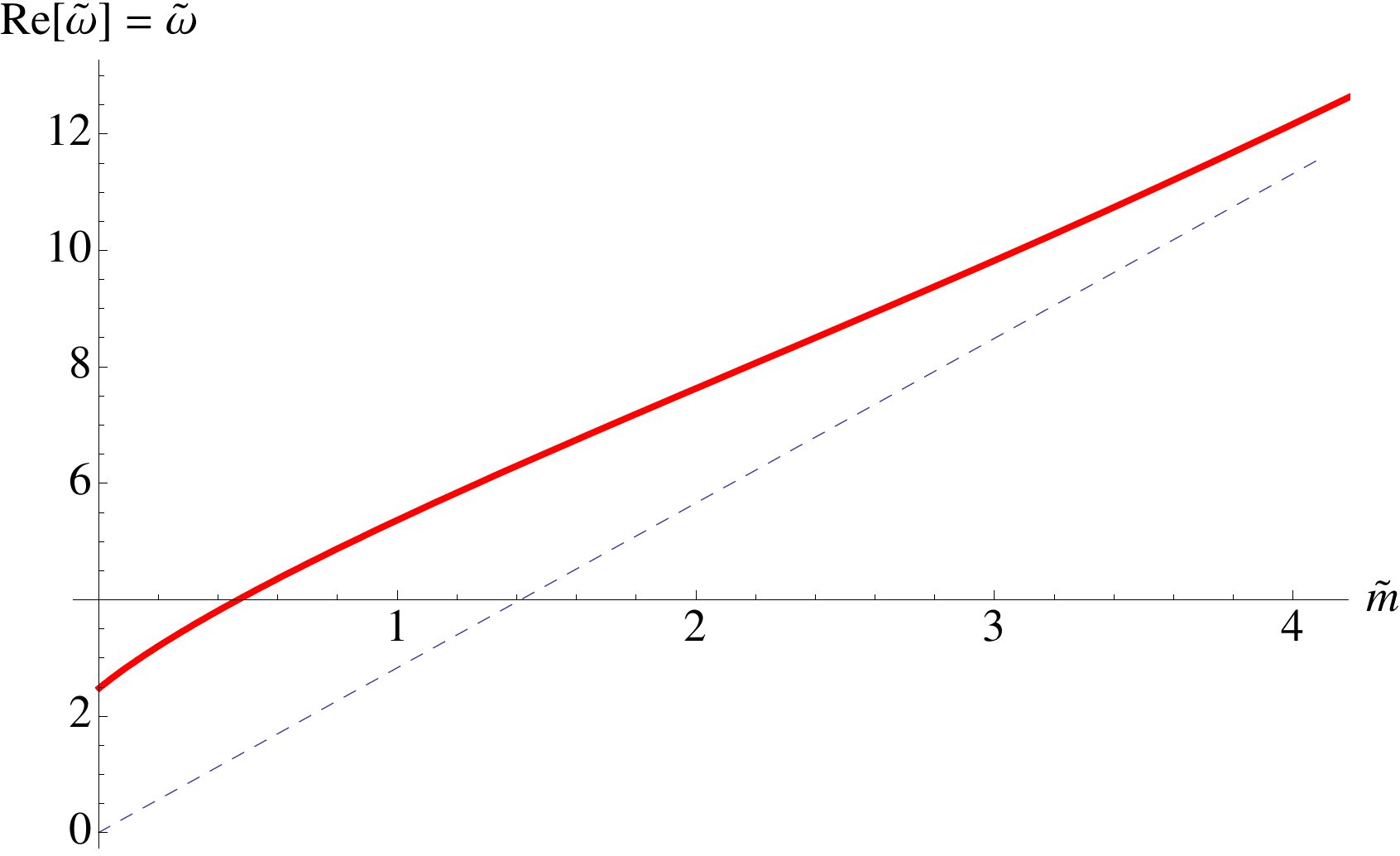}
\end{center} 
\caption{\small The $\chi$ meson mass as a function of bare quark mass for $\eta = 10$.  The dashed (black) line corresponds to the pure AdS$_5 \times S^5$ solution.}
\label{fig:fluctuations in theta eta10}
\end{figure}

In order to solve for the meson spectrum given by equation \reef{eqt:eom_theta}, we consider an ansatz for the field $\chi$ of the form:
\begin{eqnarray}
\chi = h(\tilde{z}) \exp \left(- i \tilde{\omega} t \right) \ ,
\end{eqnarray}
where we are using the same dimensionless coordinates as before, with the addition that:
\begin{equation*}\begin{array}{rclcrcl}
z &=&  b^{-2} \tilde{z} & \ , \quad & \omega &=& R^{-2} b \ \tilde{\omega} \ . \\
\end{array}
\end{equation*}
In these coordinates, the event horizon is located at $\tilde{z} = 1$.  Since there are two different types of embeddings, we analyze each case separately.  We begin by considering black hole embeddings.  In order to find the appropriate infrared initial conditions for the shooting method we use, we would like to understand the behavior of $h(\tilde{z})$ near the horizon.  The equation of motion in the limit of $\tilde{z} \to 1$ reduces to:
\begin{eqnarray} \label{eqt:quasinormal}
h''(\tilde{z}) + \frac{1}{\tilde{z}-1} h'(\tilde{z}) + \frac{\tilde{\omega}^2}{16 \left(\tilde{z}-1 \right)^2} h(\tilde{z}) &=& 0 \ .
\end{eqnarray}
The equation has solutions of the form $(1-\tilde{z})^{\pm i
  \tilde{\omega} /4}$, exactly of the form of quasinormal modes
\cite{Starinets:2002br}.  Since the appropriate fluctuation modes are
in--falling modes \cite{Hoyos:2006gb}, we require only the solution of
the form $(1-\tilde{z})^{ - i \tilde{\omega} /4}$.  This is our
initial condition at the event horizon for our shooting method.  In
order to achieve this, we redefine our fields as follows:
\begin{eqnarray*}
h(\tilde{z}) &=& y(\tilde{z}) (1-\tilde{z})^{ - i \tilde{\omega} /4} \ ,
\end{eqnarray*}
which then provides us with the following initial condition:
\begin{eqnarray}
y(\tilde{z} \to 1)  &=& \epsilon \ , 
\end{eqnarray}
where $\epsilon$ is chosen to be vanishingly small in our numerical
analysis.  The boundary condition on $y'(\tilde{z} \to 1)$ is
determined from requiring the equation of motion to be regular at the
event horizon.  The solution for the fluctuation field $y(\tilde{z})$
must be comprised of only a normalizable mode, which in turn
determines the correct value for $\tilde{\omega}$.  Since we are
dealing with quasinormal modes for the black hole embeddings,
$\tilde{\omega}$ will be complex; the real part of $\tilde{\omega}$
corresponds to the mass of the meson before it melts, and the
imaginary part of $\tilde{\omega}$ is the inverse lifetime (to
a factor of 2) \cite{Hoyos:2006gb}.
We begin by considering the trivial embedding $\theta_0(\tilde{z}) =
0$ (a black hole embedding).  This embedding corresponds to having a
zero bare quark mass.  The equation of motion \reef{eqt:eom_theta}
simplifies tremendously in this case:
\begin{eqnarray}
h''(\tilde{z}) + \left( \frac{2 \tilde{z}}{\tilde{z}^2 - 1} - \frac{1}{\tilde{z} \left(1+ \tilde{z}^2 \eta^2 \right)} \right) h'(\tilde{z}) + \frac{3 + \tilde{z} \left(-3 \tilde{z} + \tilde{\omega}^2 \right)}{ 4 \tilde{z}^2 \left(\tilde{z}^2 - 1\right)^2} h(\tilde{z}) &=& 0\ .
\end{eqnarray}
We show solutions for $\tilde{\omega}$ in figure~\ref{fig:trivial
  fluctuations in theta} as a function of the magnetic field $\eta$.
In particular, we find the same additional mode discussed in
ref.~\cite{Filev:2007qu}.  This mode becomes massless and eventually
tachyonic at approximately $\eta \approx 9.24$.  This point was
originally presented in figure~\ref{fig:ccrit}, where the intermediate
unstable phase joins the trivial embedding.  This is exactly when the
$-\tilde{c}$ vs $\tilde{m}$ plot has negative slope for all black hole
embeddings.
\begin{figure}[ht] 
\begin{center}
\includegraphics[width=10cm]{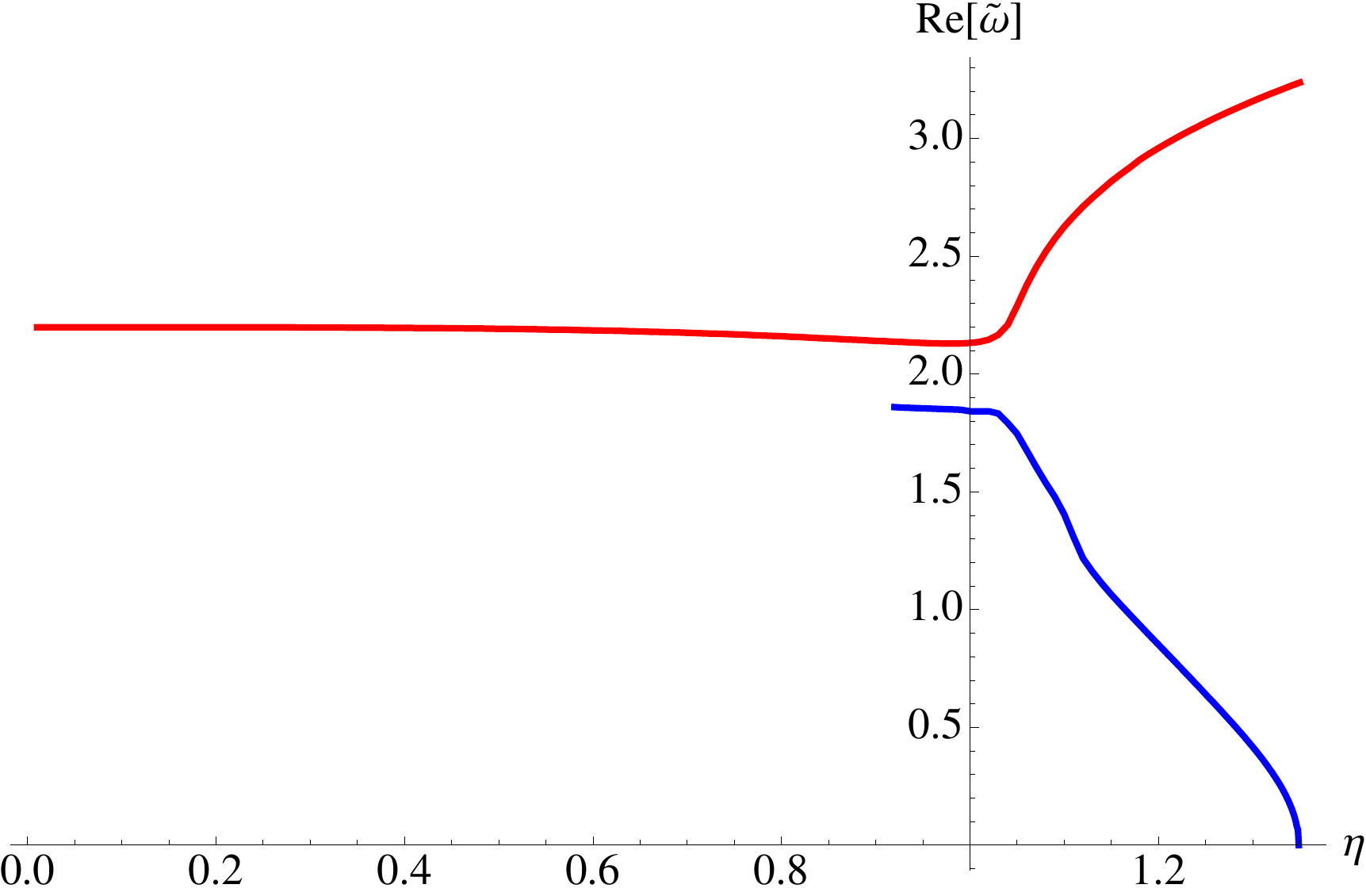}
\end{center}
\caption{\small The $\chi$ meson mass as a function of magnetic field for the trivial embedding.  The upper (red) curve is the generalization of the mode described in ref.~\cite{Hoyos:2006gb}.  The lower (blue) curve is the generalization the mode discussed in ref.~\cite{Filev:2007qu}.  We do not extend this second curve to small $\eta$ because the numerics become unreliable.}  \label{fig:trivial fluctuations in theta}
\end{figure}
\\
We now consider embeddings with non--zero bare quark mass.  This means
solving the full equation \reef{eqt:eom_theta}.  We have both
embeddings to consider; for the black hole embeddings, we will follow
the same procedure presented above to solve for the complex
$\tilde{\omega}$.  We can still use the same procedure because in the
limit of $\tilde{z} \to 1$, the equation of motion still reduces to
equation \reef{eqt:quasinormal}.  For the Minkowski embeddings, we do
not have quasinormal modes, and $\tilde{\omega}$ is purely real.
Therefore, we use as initial conditions:
\begin{eqnarray}
\chi(\tilde{z} \to \tilde{z}_{\mathrm{max}})  &=& \epsilon \ , \\
\chi'(\tilde{z} \to \tilde{z}_{\mathrm{max}}) &=& \infty \ . 
\end{eqnarray}
In figures \ref{fig:fluctuations in theta eta1} and
\ref{fig:fluctuations in theta eta10}, we show solutions for very
different magnetic field values.  In the former case, $\eta$ is small,
and we do not have chiral symmetry breaking; in the latter, $\eta$ is
large, and we have chiral symmetry breaking.  It is important to note
that in neither of the graphs do we find a massless mode at zero bare
quark mass.  In figure \ref{fig:fluctuations in theta eta1}, we find
that fluctuations about both the black hole and Minkowski embeddings
become massless and tachyonic (we do not show this in the graph).  The
tachyonic phase corresponds exactly to the regions in the $-\tilde{c}$
vs $\tilde{m}$ plot with negative slope.
\\
\section{The $\Phi$ and $A$ Meson Spectra}
\begin{figure}[ht] 
\begin{center}
\includegraphics[width=10cm]{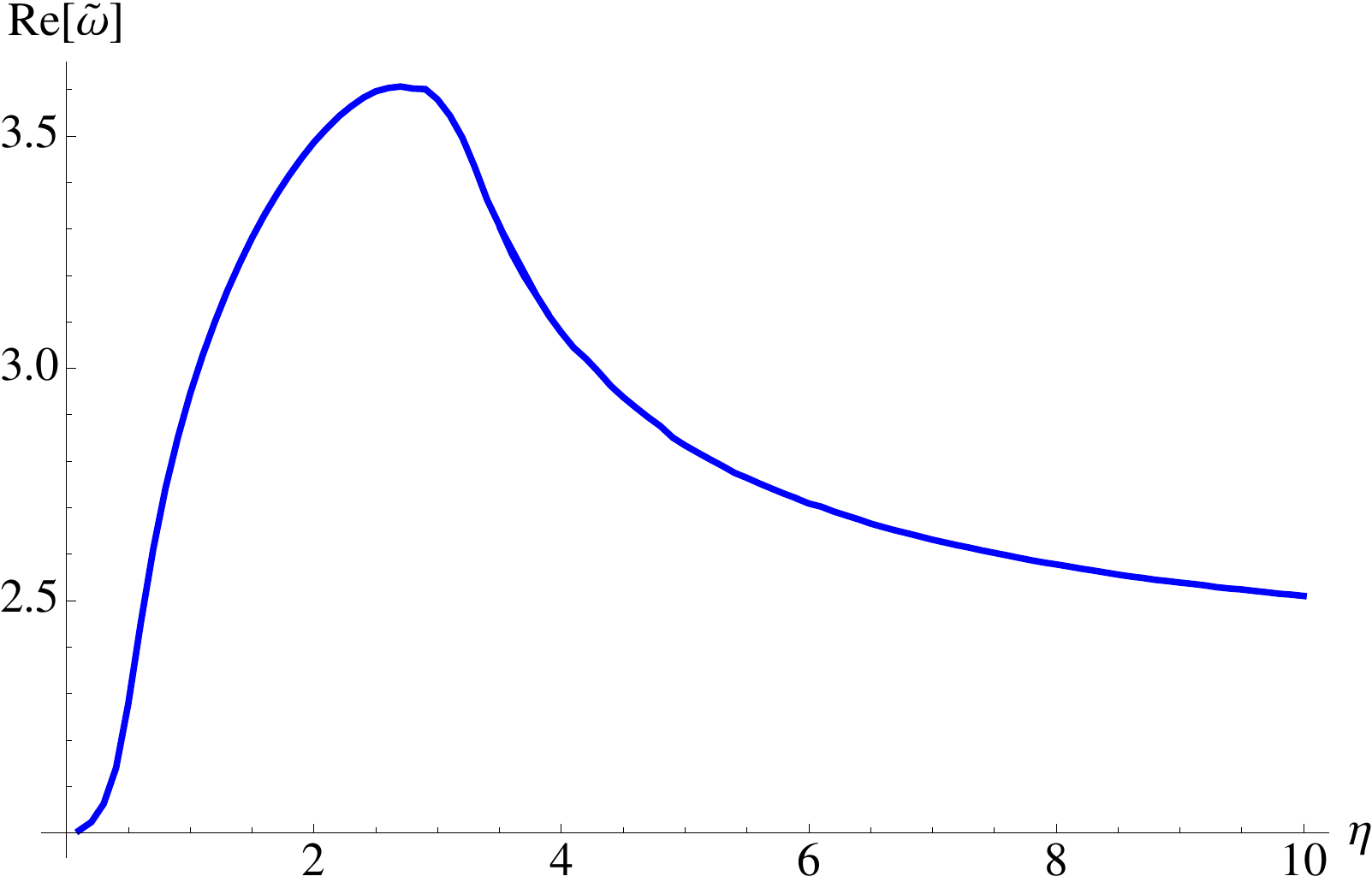}
\end{center}
\caption{\small The $A$ meson mass as a function of magnetic field for the trivial embedding.}  \label{fig:trivial fluctuations in A}
\end{figure}

\begin{figure}[ht] 
\begin{center}
\subfigure[Mass] {\includegraphics[angle=0,
width=3in ]{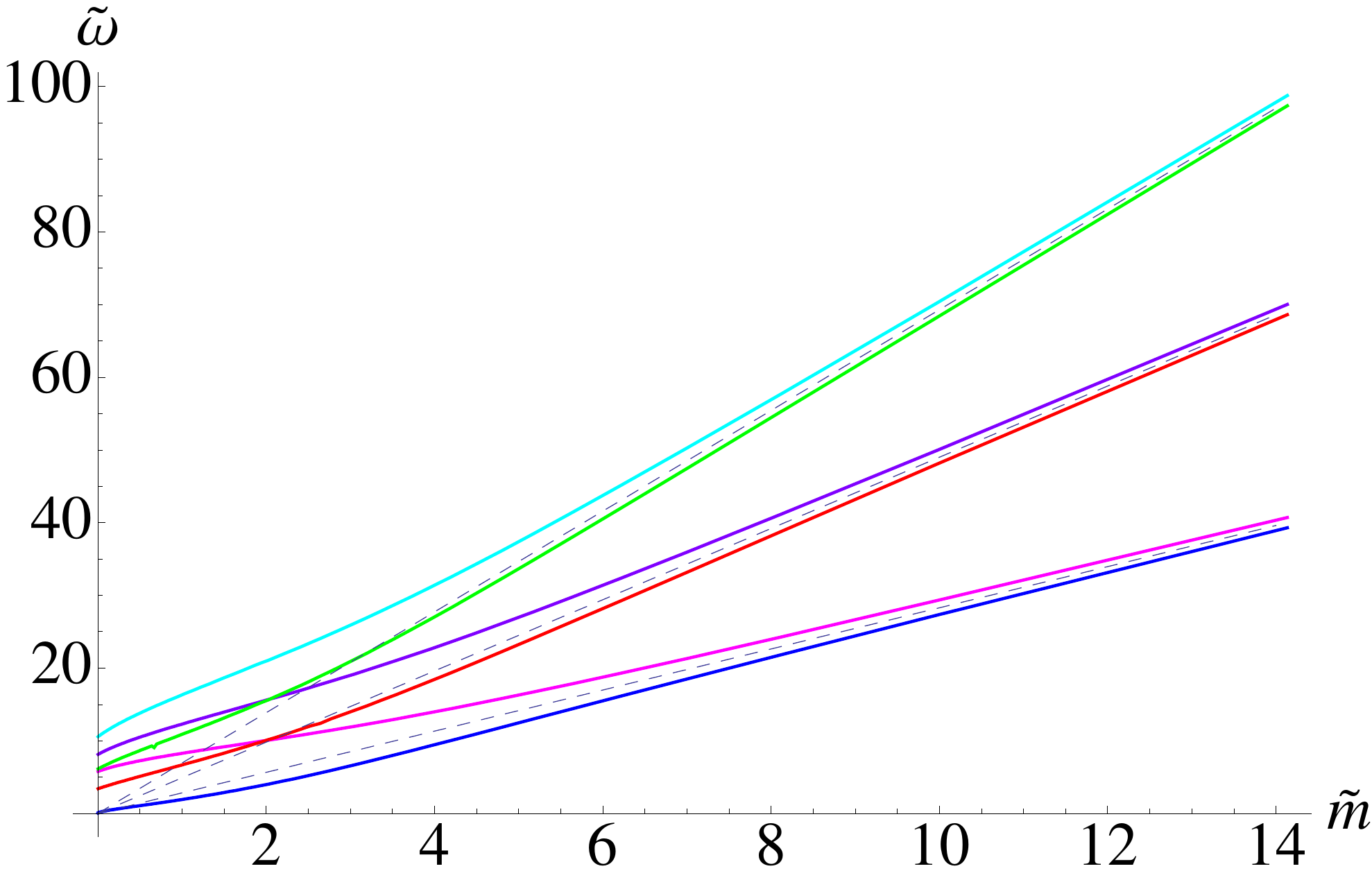}} \hspace{1cm}
\subfigure[Zoom near zero bare quark mass for the lowest mode] {\includegraphics[angle=0,
width=3in]{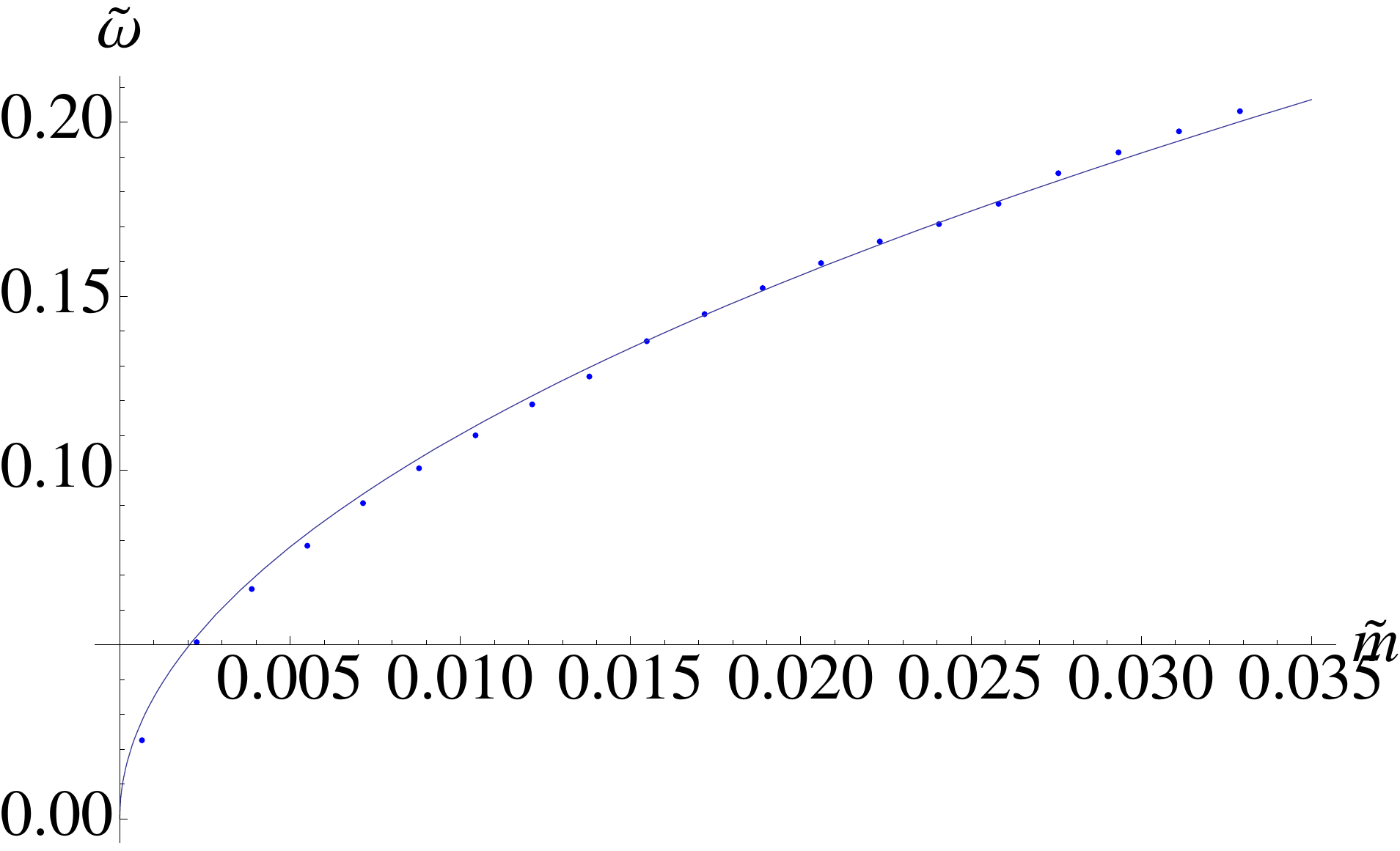}}
\end{center}
\caption{\small Coupled $A-\phi$ fluctuations for $\eta = 10$.   The dashed (black) line corresponds to the pure AdS$_5 \times S^5$ solution.}  \label{fig:fluctuations in A-phi}
\end{figure}

Let us now consider the coupled fluctuations of $\Phi$ and $A$ in
equations \reef{eqt:eom_phi_A2} and \reef{eqt:eom_phi_A1}.  We
consider an ansatz (as before) of the form:
\begin{eqnarray*}
\Phi &=& \phi(\tilde{z}) \exp \left(- i \tilde{\omega} t \right) \ , \\ 
A_1 &=& A(\tilde{z}) \exp \left(- i \tilde{\omega} t \right) \ .
\end{eqnarray*}
It is interesting to note that, for the trivial embedding
$\theta_0(\tilde{z}) = 0$, one of the coupled equations is equal to
zero, and we simply have:
\begin{eqnarray*}
A''(\tilde{z}) + \frac{\tilde{z} \left(2+\eta^2 \left( 3 \tilde{z}^2 -1 \right) \right)}{\left(\tilde{z}^2-1 \right) \left(1+ \tilde{z}^2 \eta^2 \right)} A'(\tilde{z}) + \frac{\tilde{\omega}^2}{4 \tilde{z} \left(\tilde{z}^2 - 1\right)^2 } A(\tilde{z})&=& 0 \ .
\end{eqnarray*}
Again, we note that in the limit of $\tilde{z} \to 1$, we have:
\begin{eqnarray*}
A''(\tilde{z}) + \frac{1}{\tilde{z}-1} A'(\tilde{z}) + \frac{\tilde{\omega}^2}{16 \left(\tilde{z}-1 \right)^2} A(\tilde{z}) &=& 0 \ .
\end{eqnarray*}
This is exactly the form of equation \reef{eqt:quasinormal}, so
$A(\tilde{z})$ has the same solutions of in--falling and outgoing
solutions, which provides us with the necessary initial conditions for
our shooting method.  We show solutions for $\tilde{\omega}$ in figure
\ref{fig:trivial fluctuations in A}.
Unfortunately, we do not know how to solve for the quasinormal modes
for other black hole embeddings.  Since the equations of motion are
coupled, we are unable to find an analytic solution for the
fluctuations near the event horizon.  This prevents us from using
infrared initial conditions for our shooting method.  However, we are
actually more interested in searching for the ``pion" of our system,
which will occur when we have chiral symmetry breaking.  In those
cases, we are only dealing with Minkowski embeddings with a pure real
$\omega$, so we may ignore the black hole embeddings.
\\
Since the equations of motion are coupled, it turns out that only for
specific initial conditions will both fluctuations only be comprised
of normalizable modes.  We represent this by a parameter $\alpha$
(which we must tune) as follows:
\begin{eqnarray}
A(\tilde{z} \to \tilde{z}_{\mathrm{max}}) &=& i \cos \alpha \ , \\
\phi(\tilde{z} \to \tilde{z}_{\mathrm{max}}) &=& \sin \alpha \ . 
\end{eqnarray}
The initial conditions on $A'$ and $\phi'$ are determined from the
equations of motion.  We show several solutions in figure
\ref{fig:fluctuations in A-phi}.  There are several important points
to notice.  First, we find that the lowest mode satisfies an GMOR
relationship given by:
\begin{eqnarray}
\tilde{\omega} &\approx& 1.1 \tilde{m}^{1/2} \ .
\end{eqnarray}
Therefore, we have found the Goldstone boson of our system related to
the breaking of chiral symmetry.  Second, the modes exhibit the
Zeemann splitting behavior that was discussed in
ref.~\cite{Filev:2007gb}.  It is interesting that the various modes
always cross each other at approximately $\tilde{m} \approx 2$.  We
have no intuitive explanation for this behavior.

\section{Conclusions}

We have extended the holographic study of large $N$ gauge theory in an
external magnetic field, started in ref. \cite{Filev:2007gb}, to the
case of finite temperature, allowing us to study the properties of the
quark dynamics when the theory is in the deconfined plasma phase.

The meson melting phase transition exists only below a critical value
of the applied field. This is the critical value above which
spontaneous chiral symmetry breaking is triggered (in the case of zero
mass). Above this value, regardless of the quark mass (or for fixed
quark mass, regardless of the temperature) the system remains in a
phase with a discrete spectrum of stable masses. Evidently, for these
values of the field, it is magnetically favourable for the quarks and
anti--quarks to bind together, reducing the degrees of freedom of the
system , as can be seen from our computation of the entropy.
Meanwhile, the magnetization and speed of sound are greater in this
un--melted phase.

There have been non--perturbative studies of fermionic models in
background magnetic field before, and there is a large literature (see
{\it e.g.}, the reviews of refs.~\cite{Miransky:2002eb,Wang:2007bg},
and the discussion of ref.\cite{Semenoff:1999xv} and references
therein).  Generally, those works use quite different methods to
examine aspects of the physics --- some primary non--perturbative
tools are the Dyson--Schwinger equations in various truncations). Our
results (and the zero temperature result obtained with these methods
in the zero temperature case \cite{Filev:2007gb}) are consistent with
the general expectations from those works, which is that strong
magnetic fields are generically expected to be a catalyst for
spontaneous chiral symmetry breaking in a wide class of models (see
{\it e.g.}, refs.
\cite{Gusynin:1995nb,Semenoff:1999xv,Miransky:2002eb} for a discussion
of the conjectured universality of this result).

While it is satisfying that our supergravity/string methods, which
probe the gauge theory non--perturbatively {\it via} the holographic
duality, confirm those other non--perturbative approaches, it would be
interesting and potentially useful to compare the results in more
detail, as this would (for example) allow a better understanding of
the systematics of the Dyson--Schwinger truncation schemes. Whether or
not such a direct comparison of these very different non--perturbative
approaches is possible would also be interesting to study in its own
right, potentially shedding light on other non--perturbative phenomena
in field theory that have been studied using such methods. This avenue
of investigation is beyond the scope of this paper, however, and we
leave it for future study.

\section*{Acknowledgments}
This work was supported by the US Department of Energy. This work was
presented by cvj at the Newton Institute (Cambridge, UK) conference
entitled: ``Exploring QCD: Deconfinement, Extreme Environments, and
Holography'', 20--24 August 2007. We would like to thank the
organizers for the opportunity to present and discuss our work and for
a stimulating conference. We thank Nick Dorey and Nick Evans for
drawing our attention to some of the literature on gauge theory in
external magnetic fields.

\newpage
\appendix \section{Calculating the Physical Condensate} \label{appendix:condensate}
%
The condensate (density) is given by:
\begin{eqnarray}
\langle \bar{\psi} \psi \rangle &=& \frac{\delta F}{\delta m_q} \ ,
\end{eqnarray}
where $F$ is the free energy density and $m_q$ is the physical bare
quark mass.  The free energy density is given by equation
\reef{freeenergy_gen}, which we duplicate here:
\begin{eqnarray}
F &=& 2 \pi^2 N_f T_{D7} b^4\left[\ \int\limits_{\tilde\rho_{\rm min}}^{\tilde\rho_{\rm max}}d\tilde\rho\left(\tilde\rho^3\left(1-\frac{1}{16\tilde r^8}\right)\left(1+\frac{16\eta^2\tilde r^4}{(4\tilde r^4+1)^2}\right)^{\frac12}\sqrt{1+\tilde L'^2}-\tilde\rho^3\right) \right. \nonumber \\
&& \left. - \frac{1}{4} \tilde\rho_{\rm min}^4 -\frac{1}{2}\eta^2\log\tilde\rho_{\rm max}-\frac{1}{8}\eta^2(1+\log4-\log\eta^2) \right] \nonumber \ .
\end{eqnarray}
Therefore, to calculate the condensate, we have:
\begin{eqnarray}
\langle \bar{\psi} \psi \rangle &=&  \frac{\delta F}{\delta m_q} = 2 \pi \alpha' \frac{\delta F}{\delta m} = \frac{2 \pi \alpha'}{b} \frac{\delta F}{\delta \tilde{m}} =  \frac{2 \pi \alpha'}{b} \frac{\delta F}{\delta \tilde{L}\left(\epsilon \right)} \ ,
\end{eqnarray}
where we are using a new set of coordinates $\tilde{z} = 1 /
\tilde{\rho}$ such that $\epsilon = 1 / \tilde{\rho}_{\mathrm{max}}$.
In order to continue, we consider the variation of the free energy
density:
\begin{eqnarray}
\delta F &=& 2 \pi^2 N_f T_{D7}  b^4 \left[ \ \int\limits_{\epsilon}^{\tilde{z}_{\rm max}} d \tilde{z} \left( \frac{\partial \tilde{\mathcal{L}}}{\partial \tilde{L}\left(\tilde{z}\right)}-\frac{d}{d \tilde{z}} \frac{\partial \tilde{\mathcal{L}}}{\partial \tilde{L}'\left(\tilde{z}\right)} \right)\delta \tilde{L}\left(\tilde{z}\right) + \frac{\partial \tilde{\mathcal{L}}}{\partial \tilde{L}'\left(\tilde{z}\right)} \delta \tilde{L} \left(\tilde{z}\right)\bigg|_{\epsilon}^{\tilde{z}_\mathrm{max}} \right] \ .
\end{eqnarray}
The first term is set to zero by the equation of motion.  The boundary
term evaluated at $\tilde{z}_{\mathrm{max}}$ is zero since for Minkowski
embeddings $L'(\tilde{z}_{\mathrm{max}}) = 0$ and for black hole
embeddings $ 1-{1}/{16\tilde r^8} = 0$.  Therefore, we are left
with:
%
\begin{eqnarray}
\delta F &=&- 2 \pi^2 N_f T_{D7} b^4  \frac{\partial \tilde{\mathcal{L}}}{\partial \tilde{L}'\left(\tilde{z}\right)} \delta \tilde{L}\left(\epsilon\right)\bigg|_{\tilde{z} \to \epsilon}  = - 4 \pi^2 N_f T_{D7}  b^4 \tilde{c}\delta \tilde m \ ,
\end{eqnarray}
where we have used the asymptotic expansion of $\tilde{L}(\epsilon) =
\tilde{m} + \tilde{c} z^2$.  Therefore, our final expression for the
condensate is given by:
\begin{eqnarray}
\langle \bar{\psi} \psi \rangle &=& - 8 \pi^3  \alpha' N_f T_{D7}  b^3 \tilde{c} = - \frac{T^3  \sqrt{\lambda} N_c N_f }{4} \tilde{c}\ .
\end{eqnarray}
%
\section{The $\eta$ Variation} \label{appendix:eta}
In calculating the magnetization, we need to consider the on--shell quantity:
\begin{equation}
\left(\frac{\delta \tilde I_{D7}}{\delta \eta} \right)_{T}; \quad \tilde I_{D7}=\int d\tilde\rho \ \tilde{\cal L}(\tilde{\rho}, \eta; \tilde L,\tilde L') \ .
\end{equation}
In our discussion, fixing the temperature is equivalent to fixing the
bare quark mass.  This is true because the two quantities $(T, m_q)$
are related inversely to each other by the dimensionless quantity
$\tilde{m} = m / b = 2 \pi \alpha' m_q / \pi R^2 T$.  Therefore, we
are interested in calculating:
\begin{equation}
\left(\frac{\delta \tilde I_{D7}}{\delta \eta} \right)_{\tilde m}=\int\limits_{\tilde\rho_{\rm min}}^{\tilde\rho_{\rm max}}d\tilde\rho\left(\frac{\partial\tilde{\cal L}}{\partial \tilde L'}\frac{\delta \tilde L'}{\delta\eta}+\frac{\partial\tilde{\cal L}}{\partial \tilde L}\frac{\delta \tilde L}{\delta\eta}+\frac{\partial\tilde{\cal L}}{\partial\eta}\right)=\left(\frac{\partial\tilde{\cal L}}{\partial \tilde L'}\frac{\delta \tilde L}{\delta\eta}\right)_{\tilde\rho_{\rm min}}^{\tilde\rho_{\rm max}}+\int\limits_{\tilde\rho_{\rm min}}^{\tilde\rho_{\rm max}}d\tilde\rho\frac{\partial\tilde{\cal L}}{\partial\eta}\ ,
\label{der}
\end{equation}
where we have used the equation of motion to simplify the expression.
We can further simplify the expression by noting that the boundary
term in equation \reef{der} is zero, because
\begin{equation}\frac{\partial\tilde{\cal L}}{\partial \tilde L'}\big|_{\tilde\rho_{\rm min}}=0 \ ,\quad  \mathrm{and}\quad \frac{\partial \tilde L}{\partial \eta}\big|_{\tilde\rho_{\rm max}}=\frac{\delta\tilde m}{\delta\eta}=0\ .
\end{equation}
The last relation follows from the fact that we are taking the
variational derivative with respect to $\eta$ at fixed $\tilde m$.
Therefore, the necessary computation is simplified tremendously to:
\begin{equation}
\left(\frac{\delta \tilde I_{D7}}{\delta \eta} \right)_{T} =\left(\frac{\delta \tilde I_{D7}}{\delta \eta}\right)_{\tilde m}=\int\limits_{\tilde\rho_{\rm min}}^{\tilde\rho_{\rm max}}d\tilde\rho\frac{\partial\tilde{\cal L}}{\partial\eta}\ .
\end{equation}
%
\providecommand{\href}[2]{#2}\begingroup\raggedright\endgroup

\end{document}